\documentclass[12pt]{article}%
\usepackage[nosort]{cite}
\usepackage[usenames, dvipsnames]{xcolor}
\usepackage{graphicx}
\usepackage{multicol}
\usepackage{amsfonts}
\usepackage{amssymb}
\usepackage{amsmath}
\usepackage{heck}
\usepackage{afterpage}
\usepackage{setspace}
\usepackage{verbatim}
\usepackage{longtable}
\usepackage{float}
\usepackage{subcaption}
\usepackage{epsfig}
\usepackage{enumerate}
\usepackage{epstopdf}
\usepackage[enableskew, vcentermath]{youngtab}
\usepackage{adjustbox}
\usepackage{multirow}
\usepackage{tikz}
\usepackage[margin=1in]{geometry}
\usepackage{titletoc}
\usepackage{hyperref}
\usepackage{mathtools}%
\usepackage{tikz-cd}
\providecommand{\U}[1]{\protect\rule{.1in}{.1in}}
\pdfoutput=1
\newsavebox{\mysavebox}

\hypersetup{colorlinks,citecolor=black,filecolor=black,linkcolor=black,urlcolor=black}
\usetikzlibrary{decorations.markings}

\numberwithin{equation}{section}
\def\bZ{\mathbb{Z}}

\usetikzlibrary{chains}
\allowdisplaybreaks
\tikzset{node distance=2em, ch/.style={circle,draw,on chain,inner sep=2pt},chj/.style={ch,join},every path/.style={shorten >=4pt,shorten <=4pt},line width=1pt,baseline=-1ex}

\newcommand{\ba}{\begin{eqnarray}}
\newcommand{\ea}{\end{eqnarray}}

\newcommand{\be}{\begin{equation}}
\newcommand{\ee}{\end{equation}}

\newcommand{\al}[1]{\begin{align}#1\end{align}}

\tikzstyle{startstop} = [rectangle, rounded corners, minimum width=3cm, minimum height=1cm,text centered, draw=black, fill=blue!10]
\tikzstyle{startstop} = [rectangle, rounded corners, minimum width=3cm, minimum height=1cm,text centered, draw=black, fill=blue!10]
\tikzstyle{io} = [trapezium, trapezium left angle=70, trapezium right angle=110, minimum width=3cm, minimum height=1cm, text centered, draw=black, fill=blue!30]
\tikzstyle{process} = [rectangle, minimum width=3cm, minimum height=1cm, text centered, draw=black, fill=orange!30]
\tikzstyle{decision} = [diamond, minimum width=3cm, minimum height=1cm, text centered, draw=black, fill=green!30]
\tikzstyle{arrow} = [thick,->,>=stealth]
\tikzset{->-/.style={decoration={
  markings,
  mark=at position #1 with {\arrow[scale=2.4]{>}}},postaction={decorate}}}
\makeatletter \@addtoreset{equation}{section} \makeatother

\begin{document}


\date{September 2020}

\title{S-folds, String Junctions, \\[4mm] and 4D $\mathcal{N} = 2$ SCFTs}

\institution{PENN}{\centerline{Department of Physics and Astronomy, University of Pennsylvania, Philadelphia, PA 19104, USA}}

\authors{
Jonathan J. Heckman\footnote{e-mail: {\tt jheckman@sas.upenn.edu}},
Craig Lawrie\footnote{e-mail: {\tt craig.lawrie1729@gmail.com}},
\\[4mm]
Thomas B. Rochais\footnote{e-mail: {\tt thb@sas.upenn.edu}},
Hao Y. Zhang\footnote{e-mail: {\tt zhangphy@sas.upenn.edu}},
and Gianluca Zoccarato\footnote{e-mail: {\tt gzoc@sas.upenn.edu}}}

\abstract{S-folds are a non-perturbative generalization of orientifold 3-planes which figure prominently in the
construction of 4D $\mathcal{N} = 3$ SCFTs and have also recently been used to realize examples of
4D $\mathcal{N} = 2$ SCFTs. In this paper we develop a general procedure for reading off the
flavor symmetry experienced by D3-branes probing 7-branes in the presence of an S-fold.
We develop an S-fold generalization of orientifold projection which applies to
non-perturbative string junctions. This procedure leads to a different 4D
flavor symmetry algebra depending on whether the S-fold supports discrete torsion.
We also show that this same procedure allows us to read off admissible representations of
the flavor symmetry in the associated 4D $\mathcal{N} = 2$ SCFTs. Furthermore
this provides a
prescription for how to define F-theory in the presence of S-folds with discrete torsion.
}

\maketitle

\setcounter{tocdepth}{2}

\tableofcontents


\newpage

\section{Introduction}\label{sec:INTRO}

One of the important ingredients in many string theory realizations of quantum
field theories is the use of singular geometries in the presence of various
configurations of branes. For example, in perturbative type II string theory,
all of the classical gauge groups can be realized by open strings ending on
D-branes, possibly in the presence of orientifold planes. It is also possible
to realize exceptional groups via the heterotic string, and with singular
geometries in type II / M- / F-theory compactifications.  This point of view
has led to the prediction of entirely new sorts of quantum field theories in
diverse dimensions.

As a striking example, stringy considerations led to the discovery of 4D
$\mathcal{N} = 3$ SCFTs \cite{Garcia-Etxebarria:2015wns}. These
$\mathcal{N} = 3$ theories are inherently strongly coupled, and many of them
have a realization in string theory as a stack of D3-branes on top of an
S-fold plane.\footnote{There are $\mathcal{N} = 3$ theories that come from
$\mathcal{N}=4$ super Yang--Mills with an exceptional gauge algebra which do not
have a D3-brane realization\cite{Garcia-Etxebarria:2016erx}.} The S-fold is a
generalization of the usual orientifold plane where the $\mathbb{Z}_2$
reflection symmetry is replaced by a $\mathbb{Z}_k$ symmetry, however this
only leads to a consistent supersymmetric field theory when the axio-dilaton
of Type IIB string theory is locally fixed to specific $k$-dependent values.
For additional work on $\mathcal{N} = 3$ SCFTs, see, for example, references
\cite{Garcia-Etxebarria:2015wns, Garcia-Etxebarria:2016erx,
Aharony:2015oyb,Nishinaka:2016hbw,Aharony:2016kai,Imamura:2016abe,Imamura:2016udl,Agarwal:2016rvx,
Cordova:2016emh,Lemos:2016xke,Arras:2016evy,Lawrie:2016axq,vanMuiden:2017qsh,
Amariti:2017cyd,Bourton:2018jwb,Assel:2018vtq,Tachikawa:2018njr,Ferrara:2018iko,Bonetti:2018fqz,
Garozzo:2018kra,Arai:2018utu,Garozzo:2019hbf,Arai:2019xmp,Garozzo:2019ejm,Amariti:2020lua,Zafrir:2020epd}.

Of course, rather than resorting to the full machinery of string theory one
might instead ask whether general principles of self-consistency can be used
to chart the landscape of possible quantum field theories. A notable example
of this sort of reasoning was carried out in a series of papers
\cite{Argyres:2015ffa,Argyres:2015gha,Argyres:2016xua,Argyres:2016xmc,Argyres:2016yzz,Martone:2020nsy,Argyres:2020wmq}
which established a complete classification of possible 4D $\mathcal{N} = 2$
SCFTs with a one-dimensional Coulomb branch.  A particularly interesting
feature of these results is that, at the time they were found, only some of
these theories had known string theory realizations. A key feature of this
analysis is the appearance of specific flavor symmetry algebras, as dictated
by how the Casimir invariants of the flavor symmetry translate to deformations
of the associated Seiberg--Witten curve.

Some of these 4D $\mathcal{N} = 2$ SCFTs now have known stringy realizations,
both in terms of compactifications of 6D SCFTs \cite{Ohmori:2018ona,
Giacomelli:2020jel}, as well as in terms of D3-brane probes of S-folded
7-branes \cite{Apruzzi:2020pmv}.  That being said, there are still some
theories predicted in references
\cite{Argyres:2015ffa,Argyres:2015gha,Argyres:2016xua,Argyres:2016xmc,Argyres:2016yzz,Martone:2020nsy,Argyres:2020wmq}
which have yet to be constructed.

Our aim in this paper will be to develop a general framework for understanding
the impact of S-folds on the flavor symmetries experienced by probe D3-branes
in the presence of an ambient stack of 7-branes. To this end, we develop a
prescription which generalizes the standard orientifold projection
construction for open strings, but now for more general S-folds acting on
string junctions. Doing so, we show that the structure of the resulting flavor
symmetry algebra is closely tied to the appearance of discrete torsion in the
S-fold. This is quite analogous to what happens for O3-planes, where there are
four distinct choices depending on whether a $\mathbb{Z}_2$ discrete torsion
has been activated in either the RR or NS sector. We show that the presence of
discrete torsion, in tandem with the geometric $\mathbb{Z}_k$ action on the
local geometry, leads to a well-defined set of rules which act on the
endpoints of the string junction states. This in turn leads to a general
quotienting procedure for the resulting flavor symmetry algebras. In fact, the
string junction provides more, since we can also deduce which representations
of a given flavor symmetry algebra are actually present.  For earlier work on
the use of string junctions and its relation to symmetries realized on a
7-brane, see e.g. references \cite{Gaberdiel:1997ud, DeWolfe:1998zf,
Bonora:2010bu, Grassi:2013kha, Hassler:2019eso}.  For earlier work on string
junctions in $\mathcal{N} = 3$ SCFTs, see reference \cite{Imamura:2016udl}.

The 4D $\mathcal{N} = 2$ theories that we consider will be the following. We
will start with the rank $N$ generalizations of the  Argyres--Douglas $H_0$,
$H_1$, and $H_2$ theories \cite{Argyres:1995jj}, the theory of $SU(2)$ with four fundamentals, and
the Minahan--Nemeschansky $E_6$, $E_7$, and $E_8$ theories \cite{Minahan:1996fg, Minahan:1996cj}. These theories
will be labelled as the ``parent'' theories and they are related to each other
via mass deformations from the $E_8$ Minahan--Nemeschansky theory. Furthermore
each of these parent theories has a realization as a worldvolume theory on a
stack of D3-branes in a 7-brane background (see e.g. \cite{Banks:1996nj, Noguchi:1999xq}).
We will consider the ``S-fold descendant theories'', or simply ``descendants'', as the theories obtained by
further inclusion of an S-fold plane on top of the D3-brane stack, either with or
without discrete torsion.

One of the main results of our analysis is that the resulting flavor symmetry
depends on the discrete torsion of the S-fold. In particular, we find that
when no torsion is switched on, there is a simple geometric picture available
which matches to a quotient of the associated F-theory geometry for the
7-branes. When a discrete torsion is present on the S-fold, we find that the
resulting flavor symmetry of a probe D3-brane is also different. In these
cases, the standard F-theory geometry is not valid, but we can instead deduce
its structure from the corresponding Seiberg--Witten curve of the 4D
$\mathcal{N} = 2$ SCFT.

\begin{figure}
  \centering
  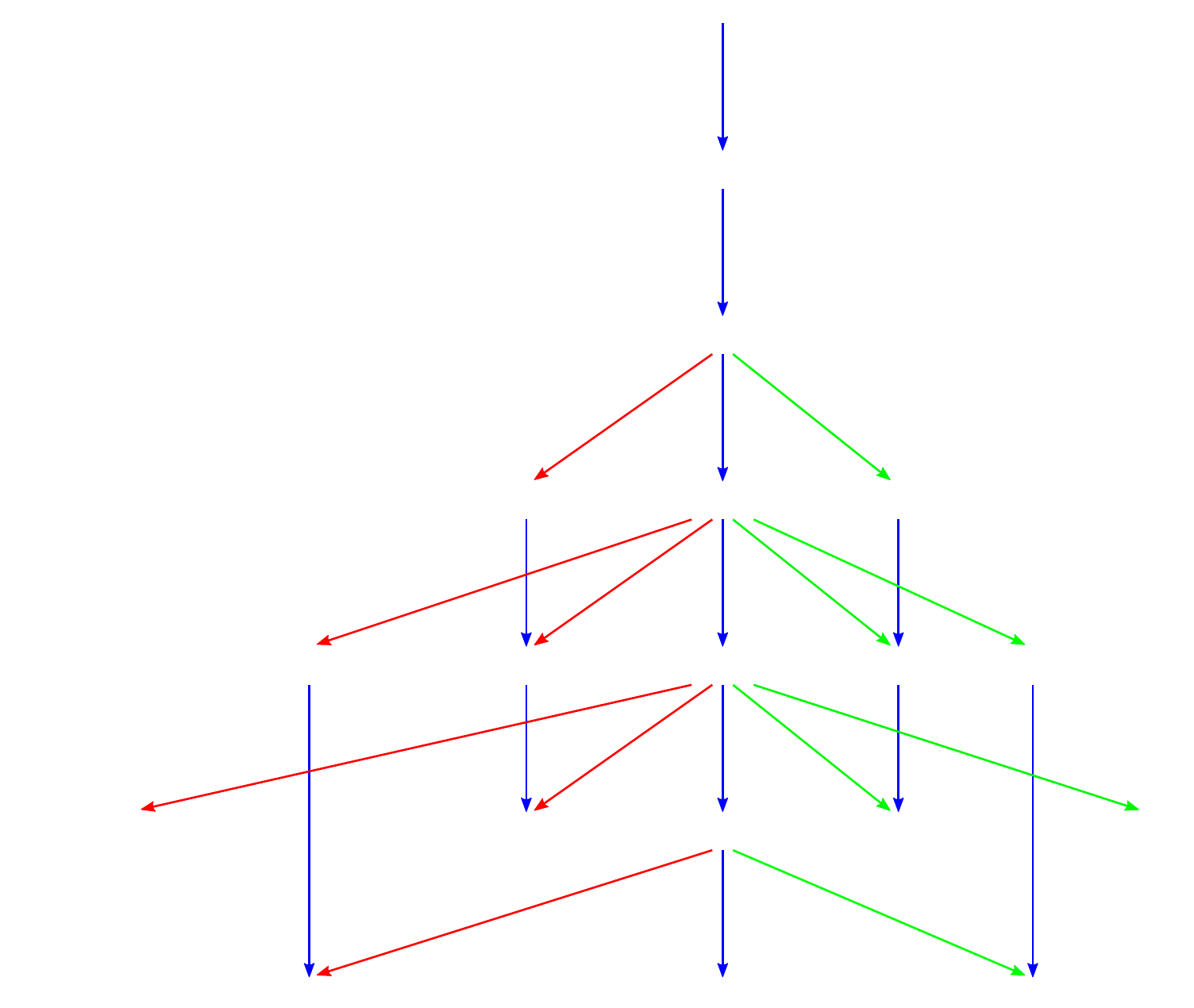
  \vspace{1em}
  \caption{Realization of the different rank one 4D $\mathcal{N} = 2$ SCFTs
  starting from the $E_8$ Minahan--Nemeschansky theory, written as $[II^{\ast}
  , E_8]$.  We can perform mass deformations (as indicated by downward blue arrows), or
  we can act by a discrete twist by an outer automorphism of an algebra,
  possibly composed with an inner automorphism.  All of the different choices
  can be realized by a suitable choice of S-fold projection with (diagonal red arrows and $\widehat{\mathbb{Z}}_{k}$)
  or without (diagonal green arrows and $\mathbb{Z}_{k}$) discrete torsion. Here, we use the conventions of
  references
  \cite{Argyres:2015ffa,Argyres:2015gha,Argyres:2016xua,Argyres:2016xmc,Argyres:2016yzz,Martone:2020nsy,Argyres:2020wmq},
  which labels a given theory by its Kodaira fiber type, as well as the
  associated flavor symmetry algebra. We note that while this notation does
  not necessarily uniquely specify a particular 4D SCFT, it does so for the
  theories listed here. The notation $\chi_a$ refers to the fact that the
  theory has a chiral deformation parameter which has scaling dimension
  $a$. The theories connected to the $[II^*, E_8]$ theory by blue arrows will be referred to as ``parent''
  theories, and the theories determined via the red/green arrows from a given
  parent will be referred to as the ``descendants'' of that parent.}\label{fig:N2rankone}
\end{figure}

Indeed, using this procedure, we show how to match each possible S-fold
quotient of 7-branes to a corresponding theory appearing in the list of rank
one 4D $\mathcal{N} = 2$ SCFTs appearing in references
\cite{Argyres:2015ffa,Argyres:2015gha,Argyres:2016xua,Argyres:2016xmc,Argyres:2016yzz,Martone:2020nsy,Argyres:2020wmq},
where the rank one theories are classified by the associated Kodaira fiber
type obtained from the Seiberg--Witten curve. In matching to our 7-brane
realization, we can visualize this process in terms of an overall quotienting
/ smoothing deformation. See table \ref{tab:AMSFOLD} for a summary of this
correspondence, and figure \ref{fig:N2rankone} for a summary of how these
different theories are related by mass deformations and discrete quotients.
Implicit in our considerations is that if we remove all the 7-branes, then we
realize $\mathcal{N} = 3$ theories, and discrete quotients thereof. An
additional comment here is that there are a few theories from
\cite{Argyres:2015ffa,Argyres:2015gha,Argyres:2016xua,Argyres:2016xmc,Argyres:2016yzz,Martone:2020nsy,Argyres:2020wmq}
which do not appear to have a simple 7-brane realization. We take this to mean
that the resulting quotients used to construct these additional theories may
not arise from purely geometric ingredients present in the ultraviolet, but
may instead involve structures which only emerge in the infrared.

\begin{table}
  \centering
  \begin{tabular}{c|c}
    Quotient & Rank One 4D $\mathcal{N} = 2$ SCFTs \cr\hline
    $IV^*/\mathbb{Z}_2$ & $[II^*, F_4]$ \cr
    $I_0^*/\mathbb{Z}_2$ & $[III^*, B_3]$ \cr
    $IV/\mathbb{Z}_2$ & $[IV^*, A_2]$ \cr
    $I_0^*/\mathbb{Z}_3$ & $[II^*, G_2]$ \cr
    $III/\mathbb{Z}_3$ & $[III^*, A_1]$ \cr
    $IV/\mathbb{Z}_4$ & $[II^*, B_1]$ \cr
    $IV^*/\widehat{\mathbb{Z}}_2$ & $[II^*, C_5]$ \cr
    $I_0^*/\widehat{\mathbb{Z}}_2$ & $[III^*, C_3C_1]$ \cr
    $IV/\widehat{\mathbb{Z}}_2$ & $[IV^*, C_2U_1]$ \cr
    $I_0^*/\widehat{\mathbb{Z}}_3$ & $[II^*, A_3 \rtimes \mathbb{Z}_2]$ \cr
    $III/\widehat{\mathbb{Z}}_3$ & $[III^*, A_1U_1 \rtimes \mathbb{Z}_2]$ \cr
    $IV/\widehat{\mathbb{Z}}_4$ & $[II^*, A_2 \rtimes \mathbb{Z}_2]$
  \end{tabular}
  \caption{For each possible discrete quotient of an F-theory
  Kodaira fiber as associated with a probe D3-brane in the presence of a 7-brane and an S-fold with or without discrete torsion,
  we find a corresponding interacting rank one theory as given in table 1 of \cite{Argyres:2016yzz}.}
  \label{tab:AMSFOLD}
\end{table}

The theories we construct include some notably subtle cases such as theories with $F_4$ flavor
symmetry. Indeed, an important point in this case is that there are some
putative 4D $\mathcal{N} = 2$ SCFTs with $F_4$ global symmetry which are now
known to be inconsistent \cite{Beem:2013sza, Shimizu:2017kzs}. These
inconsistent cases are those in which the Higgs branch of the 4D theory would
have coincided with the instanton moduli space of $F_4$ gauge theory.  Our
brane realization makes clear that we are dealing with a different theory
since in our case, we have a bulk $E_6$ 7-brane in the presence of a
codimension four S-fold with no discrete torsion. A D3-brane sitting on top of
the S-fold sees an $F_4$ flavor symmetry, while moving it inside the 7-brane
but off the S-fold results in an $E_6$ flavor symmetry.  This is also in line
with the fact that the anomalies of reference \cite{Argyres:2016yzz} are
different from the ones of the putative (and sick) $F_4$ theory ruled out in
\cite{Shimizu:2017kzs}. As an additional comment, in F-theory there are no
7-branes with 8D gauge group $F_4$, in line with the feature that such an
object does not exist either from the standpoint of F-theory, or generalized
Green--Schwarz anomalies \cite{Garcia-Etxebarria:2017crf}.

Turning the discussion around, we can also see how the emergent
Seiberg--Witten geometry for these $\mathcal{N} = 2$ theories provides an
operational definition of F-theory in S-fold backgrounds with discrete
torsion.  As a point of clarification, we note that in the single D3-brane
case there can be additional enhancements in the flavor symmetry.  The
F-theory geometry is then obtained by performing a mass deformation to the
generic flavor symmetry, and performing a further rescaling in the local
coordinates.

The rest of this paper is organized as follows. First, in section
\ref{sec:SFOLD} we present a brief review of S-folds.  In section
\ref{sec:7BRANE} we discuss the specific case of S-folds without discrete
torsion and their realization in F-theory compactifications. In section
\ref{sec:JUNCTIONS} we present a general prescription for reading off the
flavor symmetry of D3-branes probing an S-folded 7-brane.  We then use this to
provide a geometric proposal for F-theory geometry in the presence of discrete
torsion in section \ref{sec:DISCRETE}.  As a further check on our proposal, we
also compute the leading order contributions to the conformal anomalies $a$
and $c$ in the limit of a large number of probe D3-branes in section
\ref{sec:ANOMO}. Section \ref{sec:CONC} presents our conclusions. Some
additional details on brane motions in the presence of S-folds are presented
in Appendix \ref{app:BRANE}, and an explicit example of string junction projections is worked out in Appendix \ref{app:E6SJ}.

\section{S-folds} \label{sec:SFOLD}

In this section we present a brief review of S-folds. In particular, we emphasize that these objects can sometimes carry a discrete torsion.
S-fold planes are a generalization of orientifold planes introduced in
\cite{Garcia-Etxebarria:2015wns} and further studied in
\cite{Aharony:2016kai}. Initially they were used to build four dimensional
$\mathcal N=3$ supersymmetric field theories on the worldvolume of D3-branes
in the proximity of an S-fold. This was generalized in \cite{Apruzzi:2020pmv}
by adding 7-branes on top of the S-fold thus producing $\mathcal N=2$
theories. In this section we will review the construction of
\cite{Garcia-Etxebarria:2015wns} and discuss various properties of S-folds
that we shall need in the following. We will discuss the inclusion of 7-branes in section \ref{sec:7BRANE}.

\subsection{S-fold Quotients}

S-folds arise from particular terminal singularities in F-theory backgrounds
\cite{Garcia-Etxebarria:2015wns}. The singularity is produced by an orbifold
action that acts simultaneously on the base and elliptic fiber. This implies
that the geometric quotient on the base is accompanied by an $SL(2,\mathbb Z)$
action on the elliptic curve, thus explaining the name of these objects.
More concretely we consider an F-theory solution on $\mathbb
C_{(z_1,z_2,z_3)}^3 \times \mathbf T_w^2$ quotiented by a $\mathbb Z_k$ action
with generator $\sigma_k$ acting on the coordinates as
\al{ \label{eq:quot}\sigma_k : (z_1,z_2,z_3,w) \quad \rightarrow \quad (\zeta_k z_1, \zeta_k^{-1} z_2,\zeta_k z_3,\zeta_k^{-1} w)\,.
}
Here $\zeta_k$ is a $k$-th primitive root of unity. The singularity produced
is terminal as it does not admit any crepant resolution \cite{TERMINATORone,
TERMINATORtwo}.  One important observation is that in order to have a well
defined action on the torus the only allowed values of $k$ are $k=2,3,4,6$.
Compatibility with the quotient fixes the value of the complex structure
$\tau$ of the torus when $k>2$, while leaving it a free parameter for $k=2$.
The allowed values of $\tau$ as well as the $SL(2,\mathbb Z)$ action $\rho$ on
the elliptic fiber are collected in Table \ref{tab:SFOLD}. This background
preserves 12 supercharges for all values of $k>2$ and adding D3-branes probing
the singularity does not further break any additional supersymmetry (see e.g.
\cite{Becker:1996gj}). The $k=2$ case preserves 16 supercharges and therefore
produces an $\mathcal N=4$ supersymmetric theory, and the S-fold in this case
simply corresponds to the usual O3${}^{-}$-plane. Let us note that for $k = 3$ we have
chosen to use the value $\tau = \exp (2 \pi i / 3)$ which is, under a
$T$-transformation of $SL(2,\mathbb{Z})$, equivalent to taking $\exp(2 \pi i /
6)$, the ``standard'' value in the fundamental domain. This has no material
effect on any statements we make about the flavor symmetry algebra since we
can always conjugate all $SL(2,\mathbb{Z})$ generators by this
$T$-transformation anyway. The reason for this choice is to make the
$\mathbb{Z}_{k}$ action of the S-fold more manifest.

\begin{table}[!t]
\begin{center}
\begin{tabular}{c| c| c}
 & $\tau$ & $\rho$ \\[2mm]
 \hline
\rule{0pt}{1.5\normalbaselineskip} $k=2$ & free & $\left(\begin{array}{cc} -1&0\\0&-1\end{array}\right)$\\[4mm]
\hline
\rule{0pt}{1.5\normalbaselineskip}  $k=3$ & $e^{\frac{2 \pi i}{3}}$& $\left(\begin{array}{cc} -1&-1\\1&0\end{array}\right)$\\[4mm]
\hline
\rule{0pt}{1.5\normalbaselineskip}  $k=4$ & $i$& $\left(\begin{array}{cc} 0&-1\\1&0\end{array}\right)$\\[4mm]
\hline
\rule{0pt}{1.5\normalbaselineskip}  $k=6$ & $e^{\frac{2 \pi i }{6}}$& $\left(\begin{array}{cc} 0&-1\\1&1\end{array}\right)$
 \end{tabular}
 \caption{Allowed values of the Type IIB axio-dilaton $\tau$ and $SL(2,\mathbb Z)$ monodromies for various S-folds when
 no 7-branes are present.}
 \label{tab:SFOLD}
 \end{center}
 \end{table}

\subsection{Discrete Torsion}\label{sec:tor}

As in the case of orientifold 3-planes, it is possible to construct different
variants of S-folds by considering trapped three-form fluxes at the
singularity, i.e. discrete torsion. To understand the different allowed
possibilities for discrete torsion, it is helpful to consider the
asymptotic profile of the spacetime far from the singularity, as captured by a
quotient of $S^5$. As in \cite{Witten:1998xy, Aharony:2016kai}, it suffices to
consider $N$ D3-branes probing a $\mathbb Z_k$  S-fold plane. The holographic
dual in the large $N$ limit is given by Type IIB string theory on
$\text{AdS}_5 \times S^5/\mathbb Z_k$. To understand which fluxes can be
introduced it is necessary to study the cohomology of $S^5/\mathbb Z_k$, in
particular the third cohomology group which corresponds to the introduction of
three-form fluxes. In Type IIB we have two possible choices of three-form
fluxes and in the following the first component will be the NSNS flux and the
second one will be the RR flux. Usually we would simply need to compute the
cohomology with coefficients in $\mathbb Z\oplus \mathbb Z$, however due to
the fact that the S-fold action is non-trivial on the fluxes it is necessary
to take cohomology with coefficients in $(\mathbb Z \oplus \mathbb Z)_\rho$
where $\rho$ is the $SL(2,\mathbb Z)$ element listed for every S-fold in table
\ref{tab:SFOLD}.  This computation was done in \cite{Aharony:2016kai} where it
was shown that $H^3(S^5/\mathbb Z_k,(\mathbb Z \oplus \mathbb Z)_\rho)$ is the
cokernel of the map $(\mathbf{id}-\rho) : \mathbb Z^2 \rightarrow \mathbb Z^2$. The
resulting cohomology groups are
\al{ H^3(S^5/\mathbb Z_2,(\mathbb Z \oplus \mathbb Z)_\rho) &= \mathbb Z_2 \oplus \mathbb Z_2\,,\\
H^3(S^5/\mathbb Z_3,(\mathbb Z \oplus \mathbb Z)_\rho) &= \mathbb Z_3\,,\\
H^3(S^5/\mathbb Z_4,(\mathbb Z \oplus \mathbb Z)_\rho) &= \mathbb Z_2\,,\\
H^3(S^5/\mathbb Z_6,(\mathbb Z \oplus \mathbb Z)_\rho) &= \mathbb \mathbb{I}.
}
The $k=2$ case reproduces the well-known example of the four different O3-planes
\cite{Witten:1998xy}. We list here all the inequivalent choices of discrete torsion for the various S-fold planes
\al{ &k =2 \,,  \qquad \left\{ (0,0),(1,0),(0,1),(1,1)\right\}\,,\\
&k =3 \,,  \qquad \left\{ (0,0),(1,0),(2,0)\right\}\,,\\
&k =4 \,,  \qquad \left\{ (0,0),(1,0)\right\}\,,\\
&k =6 \,,  \qquad \left\{ (0,0)\right\}\,.
}
One final piece of information that will be useful in the following is the D3-brane charge carried by the S-fold plane. The charge of the $\mathbb Z_k$ S-fold plane is \cite{Aharony:2016kai}
\al{ \varepsilon_{\text{D3}} = \pm\frac{1-k}{2k}\,,
}
where the plus sign refers to the case without discrete torsion and the minus sign to the case with discrete torsion.

\section{F-theory and S-folds without Torsion}\label{sec:7BRANE}

Having reviewed some basic features of S-folds, we now turn to the structure of local F-theory models
in the presence of an S-fold. Here, we study how this is detected by the worldvolume
theory of a spacetime filling D3-brane. Recall that in F-theory, the appearance of 7-branes is encoded in
the local profile of the Type IIB axio-dilaton. Strictly speaking, this geometric correspondence between the
Coulomb branch of the D3-brane moduli space and the F-theory geometry is only valid in the purely geometric phase of F-theory,
where no discrete torsion is present. Indeed, in section \ref{sec:DISCRETE}
we will later turn the discussion around and argue that
the associated Seiberg--Witten curve provides an operational \textit{definition}
of F-theory in such backgrounds.

The rest of this section is organized as follows. First, we discuss the action
of S-folds on a local Weierstrass model. These local Weierstrass models are
chosen such that they correspond to an F-theory background for the ``parent''
theories, to wit, the rank $N$ generalisations of the Argyres--Douglas,
$SU(2)$ with four flavors, and Minahan--Nemeschansky theories. After this, we turn to an explicit
analysis of the various possible S-fold quotients of such geometries,
organizing our discussion by the corresponding $\mathbb{Z}_2$, $\mathbb{Z}_3$
and $\mathbb{Z}_4$ group action.  In the case of $\mathbb{Z}_{6}$, the
admissible minimal Kodaira fibers are trivial and we get an $\mathcal{N} = 3$
theory from D3-branes probing such a singularity. Following this procedure, we
show how to recover some examples of the Seiberg--Witten geometries, and thus
physical data like the flavor symmetry algebras,
for 4D
$\mathcal{N} = 2$ SCFTs of the sort predicted in references
\cite{Argyres:2015ffa,Argyres:2015gha,Argyres:2016xua,Argyres:2016xmc,Argyres:2016yzz,Martone:2020nsy,Argyres:2020wmq}.
As a point of clarification, the flavor symmetry which is really detected in
this way is the generic one present for multiple D3-branes probing the S-fold.
There is also an $SU(2)$ flavor symmetry as associated with the rotational
group in the worldvolume of the 7-brane (but transverse to the D3-brane), and
in the special case of a single D3-brane, there can be an ``accidental''
enhancement in the infrared. In the worldvolume theory of the D3-brane, $z$
will refer to the Coulomb branch coordinate in the covering space, and $u$
will refer to the Coulomb branch coordinate in the quotient geometry. The
$M_i$ will refer to a degree $i$ Casimir invariant built from the mass
deformations of the 7-brane flavor symmetry algebra.

\subsection{Weierstrass Models}

In order to understand which kinds of 7-brane configurations are allowed in
the presence of an S-fold plane it is convenient to understand the F-theory
Weierstrass model on the orbifolded base. Specifically we consider F-theory on
the base $B=\mathbb C^3_{(z_1,z_2,z_3)}/\mathbb Z_k$ where the generator of
$\mathbb Z_k$ acts on the coordinates of the base as in \eqref{eq:quot}. For
additional details on the procedure see, for example, \cite{DelZotto:2017pti}.
The Weierstrass model on such a base is given as usual by the polynomial
\al{ y^2 = x^3 + f(z_1,z_2,z_3) x + g(z_1,z_2,z_3)\,.
}
However, due to the orbifold action on the base coordinates $f$ and $g$ become
$\mathbb Z_k$-equivariant polynomials. By the condition that the elliptic
fibration be a Calabi--Yau variety the coefficients of the Weierstrass
model, $f$ and $g$, are required to be sections of $\mathcal O(-4K_B)$ and $\mathcal
O(-6K_B)$, respectively. Homogeneity fixes $x$
to be a section of $\mathcal O(-2 K_B)$ and $y$ to be a section of $\mathcal
O(-3K_B)$. For an orbifold a section of $\mathcal O(- l K_B)$ must transform
with a factor $\text{det} (\gamma)^{l}$ where $\gamma$ is the matrix
representation of any orbifold group element acting on the coordinates. To
write down possible Weierstrass models it is convenient to expand $f$ and $g$
as polynomials in the variables $z_i$
\al{ f &= \sum_{a,b,c\geq 0} f_{abc}z_1^a z_2^b z_3^c\,,\\
g &= \sum_{a,b,c\geq 0} g_{abc}z_1^a z_2^b z_3^c\,.
}
Requiring $f$ and $g$ to transform appropriately under the orbifold action puts restrictions on the allowed polynomial coefficients $f_{abc}$ and $g_{abc}$. We list in the following the possible choices for the different S-fold planes.

\begin{itemize}
\item[-]$\underline{k = 2}$. In this case both $f$ and $g$ are invariant under the orbifold action. This fixes $f_{abc} = g_{abc}= 0 $  for $a-b+c \neq 0 \mod 2$. The lowest order terms are the constant ones giving generically a smooth elliptic curve with constant complex structure over $\mathbb C^3$.\footnote{Note that this does not mean that the orbifold action is trivial on the elliptic curve. Indeed the coordinate $y$ changes sign under the action of the generator of $\mathbb Z_2$.}

\item[-]$\underline{k = 3}$. In this case the orbifold action implies that $g$ is invariant and $f \rightarrow e^{2 \pi i/3 } f$. This fixes $f_{abc} = 0$ for $a-b+c \neq 1 \mod 3$ and $g_{abc} = 0 $ for $a-b+c \neq 0 \mod 3$. 
%
%

\item[-]$\underline{k = 4}$. In this case the orbifold action implies that $f$ is invariant and $g \rightarrow - g$. This fixes $f_{abc} = 0$ for $a-b+c \neq 0 \mod 4$ and $g_{abc} = 0 $ for $a-b+c \neq 2 \mod 4$. 
%
%

\item[-]$\underline{k = 6}$. In this case the orbifold action implies that $g$ is invariant and $f \rightarrow e^{4 \pi i/3 } f$. This fixes $f_{abc} = 0$ for $a-b+c \neq 4 \mod 6$ and $g_{abc} = 0 $ for $a-b+c \neq 0 \mod 6$. 
%
%

\end{itemize}

In the following we will be interested in a restricted class of Weierstrass models that preserve $\mathcal N=2$ supersymmetry. This can be achieved by taking all 7-branes to wrap the $(z_1,z_2)$-plane, implying that $f$ and $g$ will only depend on $z_3$. Moreover to simplify the notation we shall denote by $z$ the coordinate $z_3$ in the covering space.

We exclusively focus on Weierstrass models where the axio-dilaton is constant
so that we can realize an SCFT on the worldvolume of the D3-brane. F-theory
constructions with constant coupling were discussed in \cite{Dasgupta:1996ij}.
Additionally, we require that the singularity type remain minimal, which
imposes the further condition that the degrees of $f$ and $g$ as polynomials
in $z$ are $\text{deg}(f)<4 $ and $\text{deg}(g)<6$. For each possible S-fold
quotient, we list the covering space theory prior to the quotient in table
\ref{tab:weierstrass}. Note that the $k=6$ quotient does not allow any
dependence on $z$ in the Weierstrass model without incurring non-minimal
Kodaira fibers, and thus there can be no 7-branes present. This implies that
the theory will have enhanced $\mathcal N\geq 3$ supersymmetry.

A careful comparison of tables \ref{tab:SFOLD} and \ref{tab:weierstrass} also
reveals that the correlation of values of $k$ with $\tau$ are different in the presence or absence of
7-branes. This is to be expected because the presence of 7-branes impacts the profile of the axio-dilaton.

\begin{table}[!t]
\begin{center}
\begin{tabular}{c| c|c|c}
 Quotient & Weierstrass Model & Kodaira Fiber Type & $\tau$ \\
 \hline
\rule{0pt}{1.\normalbaselineskip} $k=2$ & $y^2= x^3 + z^4$ & $IV^*$ & $e^{\frac{2 \pi i}{3}}$ \\[1pt]
\hline
\rule{0pt}{1.\normalbaselineskip} $k=2$ & $y^2= x^3 + z^2 x$ & $I_0^*$ & $ i $ \\[1pt]
\hline
\rule{0pt}{1.\normalbaselineskip} $k=2$ & $y^2= x^3 + z^2 $& $IV$ & $e^{\frac{2 \pi i}{6}}$ \\[1pt]
\hline
\rule{0pt}{1.\normalbaselineskip} $k=3$ & $y^2= x^3 + z^3 $ & $I_0^*$ & $e^{\frac{2 \pi i}{6}}$ \\[1pt]
\hline
\rule{0pt}{1.\normalbaselineskip} $k=3$ & $y^2= x^3 + z x $ & $III$ & $ i $\\[1pt]
\hline
\rule{0pt}{1.\normalbaselineskip} $k=4$ & $y^2= x^3 + z^2 $ & $IV$ & $e^{\frac{2 \pi i}{6}}$ \\[1pt]
\hline
\rule{0pt}{1.\normalbaselineskip} $k=6$ & $y^2= x^3    + g_0  $ & $\varnothing$ & $e^{\frac{2 \pi i}{6}}$ \\[1pt]
\hline
 \end{tabular}
 \caption{Allowed values of S-fold projection compatible with a specified minimal Kodaira fiber type.
 Here we drop all higher order singularities and focus on the specific situation where the axio-dilaton is constant.}
 \label{tab:weierstrass}
 \end{center}
 \end{table}

The relevance of the Weierstrass model is that it will allow us to read off
the Seiberg--Witten curve of the resulting $\mathcal N=2$ theory for the case
of a single D3-brane probe. Indeed in this case the Seiberg--Witten curve can be identified with
the elliptic fiber of the F-theory model and the coordinate $z$ becomes the Coulomb branch parameter of
the theory. In the following we will discuss each possible case leading to a
rank one SCFT writing down the Seiberg--Witten curve and match the results to
the ones known in the literature. We would like to stress that the procedure works only in the case \emph{without} discrete torsion, and in the presence of discrete torsion we do not have a procedure to read off the Seiberg--Witten curve from the geometry. We will confirm the various identifications via a string junction analysis in section \ref{sec:JUNCTIONS} where we will also be able to identify the theories on the probe D3-branes also in the presence of discrete torsion.
Before turning to the discussion of each case separately we would like to point out that in the above
we have been using the covering space coordinates. It is also helpful to work directly in terms of a
local coordinate in the quotient geometry. In general for a $\mathbb Z_k$ quotient we would need to use $u = z^k$ which is invariant under the quotient. To find the appropriate invariant combinations for $x$ and $y$ we can use the fact that under the general rescaling \cite{0907.0298,Argyres:2016yzz,Apruzzi:2020pmv}
 \al{ x \mapsto \lambda^2 x\,, \quad y \mapsto \lambda^3 y\,,
 }
 which modifies $f$ and $g$ as
 \al{f \mapsto \lambda^{-4} f \,, \quad g \mapsto \lambda^{-6} g\,,
 }
the elliptic fibration is left invariant. By choosing $\lambda = z^{1-k}$ the rescaled $x$ and $y$ variables will be invariant under the $\mathbb Z_k$ quotient.

Using this information we will be able to write down the Seiberg--Witten curves for the various rank one theories.

\subsection{$\mathbb Z_2$ Quotients }

In this subsection we turn to $\mathbb{Z}_2$ quotients of an F-theory model. This sort of quotient
can be taken for parent theories with an $E_6$ 7-brane, as realized by a type $IV^{\ast}$ fiber,
a $D_4$ 7-brane, as realized by a type $I_{0}^{\ast}$ fiber, and an $H_2$ 7-brane, as realized by
a type $IV$ fiber.

\subsubsection{Quotient of $E_6$}

The Weierstrass model for an $E_6$ singularity can be written as
\al{ y^2 = x^3 + z^4\,.
}
Homogeneity fixes the scaling dimension of $z$ to be $\Delta(z) = 3$. The
maximal deformation of the singularity compatible with the $\mathbb Z_2$
quotient involves introducing the following $M_i$:
\al{y^2= x^3 +x \left(M_8 + M_2 z^2 \right)+ z^4 +  M_6 z^2 +M_{12}\,.
}
Here we chose the convention to label the mass deformations of the 4D
$\mathcal{N} = 2$ SCFT as degree $i$ Casimir invariants $M_i$ where the
scaling dimension is $\Delta(M_i) = i$. We can now move to the quotient space
by performing the aforementioned rescaling. Let us be explicit in this first
case. The scaling is
\begin{equation}
  x \rightarrow z^{-2} x \,, \quad y \rightarrow z^{-3} y \,,
\end{equation}
which leads to an overall factor on the $y^2$ and $x^3$ terms in the Weierstrass
equation of $z^{-6}$. Removing this denominator is equivalent to the rescaling
\begin{equation}
  f \rightarrow z^4 f \,, \quad g \rightarrow z^6 g \,,
\end{equation}
as described in the general case in \cite{0907.0298}. After this rescaling
we perform the replacement with the quotiented coordinate, $u$, via $u
= z^2$. The resulting model becomes
\al{y^2= x^3 +x \left(M_8 u^2 + M_2 u^3 \right)+ u^5 +  M_6 u^4 +M_{12} u^3\,,
}
where we have used the same notation $x$ and $y$ for before and after the
rescaling. In this case turning
off all mass deformations we obtain a $II^*$ singular fiber at the origin.
Comparing with \cite{Argyres:2015gha} we see that this Weierstrass model
matches the Seiberg--Witten curve of the
$[II^*,F_4]$ theory.

\subsubsection{Quotient of $D_4$}

The $D_4$ singularity admits two different minimal Weierstrass presentations,
one of which is compatible with the $\mathbb{Z}_2$ quotient and the other
which is compatible with the $\mathbb{Z}_3$ quotient.
For the $\mathbb Z_2$ quotient we have the Weierstrass model
\al{ y ^2 = x^3 + x z^2\,.
}
Homogeneity fixes the scaling dimension of $z$ to be $\Delta(z) = 2$, and the
deformation of the singularity compatible with the $\mathbb Z_2$ quotient is
given by the introduction of the Casimirs $M_2$, $M_4$, and $M_6$:
\al{y^2= x^3 +x \left(M_4 + z^2 \right)+  M_2 z^2 +M_{6}\,.
}
Again we move to the quotient space by performing the rescaling,
as described above. After rescaling the model becomes
\al{y^2= x^3 +x \left(M_4 u^2 +  u^3 \right)+   M_2 u^4 +M_{6} u^3\,.
}
In this case turning off all mass deformations we obtain a $III^*$ singular
fiber at the origin, and if we compare with \cite{Argyres:2015gha} we see that this
Weierstrass model matches the Seiberg--Witten curve of the $[III^*,B_3]$
theory listed therein.

\subsubsection{Quotient of $H_2$}

The Weierstrass model for an $H_2$ singularity, also known as a type $IV$
fiber, is
\al{ y ^2 = x^3 + z^2\,.
}
As usual the scaling dimension of $z$ is fixed by homogeneity of the
Weierstrass equation. We have $\Delta(z) =3/2$. The singularity can be
deformed in such a way that is compatible with a $\mathbb{Z}_2$ quotient by
introducing $M_2$ and $M_3$ as follows:
\al{y^2= x^3 +x M_2 +  z^2 +M_{3}\,.
}
The resulting model in the quotient space is obtained by performing the
now-familiar rescaling:
\al{y^2= x^3 +x M_2 u^2 +   u^4 +M_{3} u^3\,.
}
We can see that turning off all mass deformations we obtain a $IV^*$ singular
fiber at the origin. Comparing with \cite{Argyres:2015gha} we see that this
Weierstrass model is precisely giving the Seiberg--Witten curve of the $[IV^*,A_2]$
theory.

\subsection{$\mathbb Z_3$ Quotients }

We next turn to $\mathbb{Z}_{3}$ quotients of a local F-theory geometry. This
can be carried out for a $D_4$ 7-brane, as realized by a type $I_{0}^{\ast}$
fiber, and an $H_1$ 7-brane, as realized by a type $III$ fiber.

\subsubsection{Quotient of $D_4$}

The other Weierstrass model for the $I_0^*$ singularity, the one compatible with the $\mathbb Z_3$ symmetry, is:
\al{ y ^2 = x^3 + z^3\,,
}
and homogeneity fixes the scaling dimension of $z$ to be $\Delta(z) =2$. The
deformation of the singularity compatible with the $\mathbb Z_3$ quotient is
\al{y^2= x^3 +M_2 x  z  + M_6 + z^3\,.
}
We can now move to the quotient space by performing the aforementioned rescaling. The resulting model becomes
\al{y^2= x^3 +x M_2 u^3 +   u^5 +M_{6} u^4\,.
}
In this case turning off all mass deformations we obtain a $II^*$ singular
fiber at the origin. Comparing with \cite{Argyres:2015gha} we see that this
Weierstrass model matches the Seiberg--Witten curve of the $[II^*,G_2]$
theory.

\subsubsection{Quotient of $H_1$}

The Weierstrass model for an $H_1$ singularity, or type $III$ fiber, compatible with the $\mathbb Z_3$ symmetry is
\al{ y ^2 = x^3 + x z\,.
}
Homogeneity fixes the scaling dimension of $z$ to be $\Delta(z) =4/3$. The deformation of the singularity compatible with the $\mathbb Z_3$ quotient is
\al{y^2= x^3 + x  z  + M_2 \,.
}
As usual we can move to the quotient space by performing the rescaling
described above. The resulting model becomes
\al{y^2= x^3 +x  u^3 +   M_{2} u^4\,.
}
In this case turning off all mass deformations we obtain a $III^*$ singular
fiber at the origin, and a comparison with \cite{Argyres:2015gha} shows that
this Weierstrass model reproduces the Seiberg--Witten curve of the
$[III^*,A_1]$ theory.

\subsection{$\mathbb Z_4$ Quotient of $H_2$}

Finally, we turn to the case of $\mathbb{Z}_{4}$ quotients. In this case there is only a single
choice available, as given by an $H_2$ 7-brane, namely a type $IV$ fiber.
Recall that the Weierstrass model for an $H_2$ singularity is
\al{ y ^2 = x^3 + z^2\,,
}
and that homogeneity of the polynomial fixes the scaling dimension of $z$ to
be $\Delta(z) =3/2$. The deformation of the singularity compatible with the
$\mathbb Z_4$ quotient allows the introduction of only a single Casimir invariant $M_2$:
\al{y^2= x^3 + M_{2} x    + z^2 \,.
}
We can pass to the quotient space geometry by performing the aforementioned rescaling.
The resulting Weierstrass model is
\al{y^2= x^3 +M_2 x  u^3 +    u^5\,.
}
Turning off all mass deformations we obtain a $II^*$ singular fiber at the
origin. Comparing with \cite{Argyres:2015gha} we see that this Weierstrass
model matches the Seiberg--Witten curve of the $[II^*,B_1]$ theory.

\section{String Junctions}\label{sec:JUNCTIONS}

In the previous section we presented a general analysis of how to read off the Seiberg--Witten curve for the worldvolume theory
of a probe D3-brane in the presence of a 7-brane and an S-fold without discrete torsion.
Geometrically, this provides a satisfying picture for how to realize a subset of possible 4D $\mathcal{N} = 2$ SCFTs,
but it also leaves open the question as to whether we can also understand quotients with discrete torsion. An additional issue
is that in all cases the information of the flavor symmetry is encoded indirectly in the Seiberg--Witten curve via the unfolding of the singularity.

To provide a systematic analysis of cases with and without discrete torsion, we now analyze the spectrum of string junctions
in the presence of an S-fold. The rules we develop lead to a different quotienting procedure for the flavor symmetry algebra,
and the available options are all contained in the options predicted in references
\cite{Argyres:2015ffa,Argyres:2015gha,Argyres:2016xua,Argyres:2016xmc,Argyres:2016yzz,Martone:2020nsy,Argyres:2020wmq}.
Again, we must add the caveat that our analysis really leads to a derivation of the generic flavor symmetry, namely the one which is present for
multiple probe D3-branes.

To better understand how S-fold projection works, we first review the standard
case of orientifold projection for oriented perturbative strings, we follow
this with the rules for S-fold projection in the case of $\mathbb{Z}_2$,
$\mathbb{Z}_3$, and $\mathbb{Z}_4$ quotients. We then turn to the explicit
S-fold projections for string junctions attached to 7-branes.

In what follows, we will find it useful to arrange the bound states of
$[p,q]$ 7-branes so that the group action amounts to a simple rearrangement
operation on these stacks. We refer to these branes according to the resulting
$SL(2,\mathbb{Z})$ monodromy on the axio-dilaton, writing the monodromy as:
\begin{equation} \label{monomove}
M_{[p,q]} = \left[
\begin{array}
[c]{cc}%
1 + pq & - p^2 \\
q^2 & 1 - pq
\end{array}
\right] \,,
\end{equation}
for a $[p,q]$ 7-brane. We will frequently refer to the branes:
\begin{align}
A & = M_{[1,0]} = \left[
\begin{array}
[c]{cc}%
1 & -1\\
0 & 1
\end{array}
\right]  \text{, \ \ }B= M_{[1,-1]} = \left[
\begin{array}
[c]{cc}%
0 & -1\\
1 & 2
\end{array}
\right]  \text{, \ \ }
C= M_{[1,1]} = \left[
\begin{array}
[c]{cc}%
2 & -1\\
1 & 0
\end{array}
\right]  \text{, \ \ } \nonumber \\
D & = M_{[0,1]} = \left[
\begin{array}
[c]{cc}%
1 & 0\\
1 & 1
\end{array}
\right]  \text{, \ \ }
X= M_{[2,-1]} = \left[
\begin{array}
[c]{cc}%
-1 & -4\\
1 & 3
\end{array}
\right]  \text{, \ \ }
Y= M_{[2,1]} = \left[
\begin{array}
[c]{cc}%
3 & -4\\
1 & -1
\end{array}
\right] \,.
\end{align}
We will also need to rearrange our branes to make the S-fold quotient more manifest.
We accomplish this with different brane arrangements (see Appendix \ref{app:BRANE}). This includes:
\begin{align}
E_{6} &: A^{5} B C^2 \sim A^{6} XC \sim AAA C AAA C \\
D_{4} &: A^{4} BC \sim AA C AA C \sim AABBDD \\
H_{2} &: A^{3} C \sim AC AC \sim AYAY \sim DADA \\
H_{1} &: A^{2} C \sim ABD \,.
\end{align}
These 7-branes correspond to the
F-theory backgrounds that give rise to the parent theories on the probe
D3-branes, when there is no S-fold. We again stress that the symmetry algebra
obtained when we include the S-fold is the one enjoyed by the probe D3-branes.

The utility in introducing these different brane systems is that we can then read off the corresponding root system as well as representations from string junctions stretched between these different constituent branes. As a point of notation, we write $a_{i}$ to denote weights associated with $A$-branes, with similar conventions for the $B$, $C$, and $D$ branes, and where the presence of a minus sign indicates the orientation of
the string. For example, the roots of $SU(N)$ for a stack $A^{N}$ would then be represented as $(a_i - a_j)$ for $i,j = 1,...,N$ and $i \neq j$. A junction with endpoints on different types of branes is represented similarly by an oriented graph with weights. Elements of the Cartan subalgebra correspond to string junctions which begin and end on the same branes.

\subsection{Orientifold Projection}

\begin{figure}[t!]
\begin{center}
\includegraphics[scale=.8]{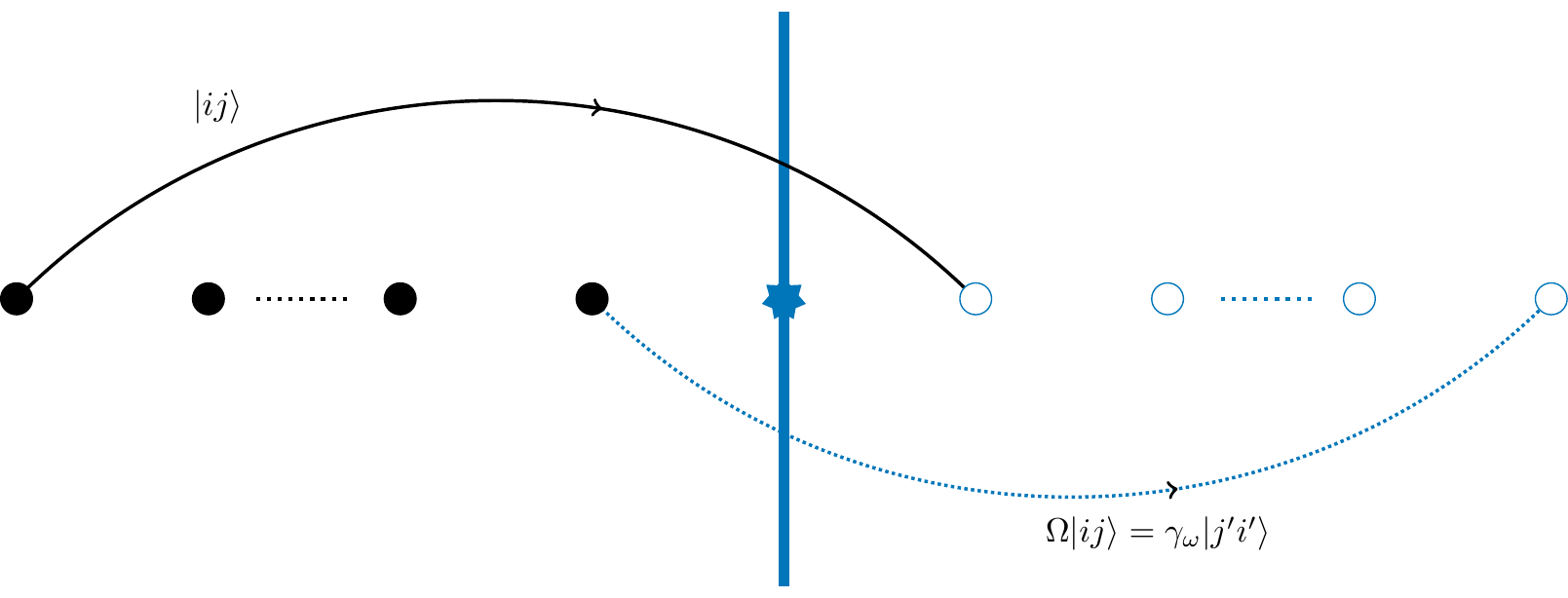}
\end{center}
\caption{Illustration of orientifold projection acting on perturbative open strings. We denote the orientifold image branes by
open shapes, and image strings by dashed blue lines.}
\label{fig:oplane}
\end{figure}

Before delving into how the S-fold planes act on the string junctions stretching between 7-branes, we first review how the usual orientifold planes that appear in perturbative string theory act on string states. Recall that in the presence of a stack of $2N$ D-branes, open string states containing a vector are labelled by Chan--Paton factors $\lambda_{ij}$ for $i,j = 1,\dots, 2N$. Each $\lambda_{ij}$ state is an open string stretching between the $i$-th and the $j$-th brane. When the stack of D-branes sits on top of an orientifold plane it is necessary to specify the action of worldsheet orientation reversal on these states. The general action is
\al{ \Omega : \lambda \ \mapsto\  - M \lambda^T M^{-1}\,.
}
The minus sign appears because of the effect of worldsheet parity on the open
string oscillators and transposition appears because the endpoints of an open
string are interchanged. $M$ is an additional conjugation on the endpoints and
consistency fixes it to be either symmetric or anti-symmetric. When $M$ is
chosen to be symmetric the resulting Lie algebra on the stack of D-branes will
be $D_N$ and when $M$ is anti-symmetric the Lie algebra will be $C_N$. Given
this we will label the symmetric choice $M_{\text{SO}}$ and the anti-symmetric
one $M_{\text{Sp}}$. In the following we will choose\footnote{Note that it is
customary in the literature to choose $M_{\text{SO}}$ to be the identity
matrix. Our choice will give isomorphic algebras after projection and leads to
a simpler geometric picture in terms of branes probing the orientifold plane.}
\al{ M_{\text{SO}} &= \left(\begin{array}{cc}0 & J_N \\ J_N &0\end{array}\right)\,,\\
 M_{\text{Sp}} &= \left(\begin{array}{cc}0 & iJ_N \\ -iJ_N &0\end{array}\right)\,.
}
Here $J_N = \delta_{i+j,N+1}$ with $i,j = 1,\dots, N$, namely the anti-diagonal matrix in which non-zero entries are all equal to 1.
We can therefore explicitly write the action of $\Omega$ on the various string states which we label as $|ij\rangle$ for a string stretching between the $i$-th and $j$-th brane. We will find it convenient to use the notation  $i' = 2N+1-i$. The map is:
\al{ \Omega |ij\rangle = \gamma_{\Omega}|j' i'\rangle\,.
}
Here, the choice of phase factor is specified via (see figure \ref{fig:oplane}):
\begin{align}
\text{Sp projection} &\rightarrow \gamma_\Omega = 1\\
  \text{SO projection} &\rightarrow \left\{\begin{array}{ll}\gamma_\Omega = -1, & \text{string crosses orientifold} \\\gamma_\Omega=-1,& i=j\\ \gamma_\Omega = 1, &\text{otherwise.} \end{array}\right.
\end{align}

Finally, it is important to understand which projection corresponds to which
orientifold plane. The system that more closely resembles the ones we will
study in the following is a stack of D7-branes on top of an orientifold
3-plane. Recall that there exist four different orientifold planes usually
called $\text{O3}^-$, $\widetilde{\text{O3}}^-$, $\text{O3}^+$,  and
$\widetilde{\text{O3}}^+$. In terms of the discrete torsion introduced in
section \ref{sec:tor} they have torsion $(0,0)$, $(0,1)$, $(1,0)$ and $(1,1)$
respectively. The first two give a $D$-type algebra on a stack of D3-branes
and a $C$-type algebra on a stack of D7-branes, the last two give a $C$-type
algebra on a stack of D3-branes and a $D$-type algebra on a stack of
D7-branes. The action on other kinds of 7-branes can be obtained via
$SL(2,\mathbb Z)$ conjugation knowing that the $\text{O3}^-$ plane is
invariant under $SL(2,\mathbb Z)$ and that the action of $SL(2,\mathbb Z)$ for
the other planes can be inferred by looking at the action on the plane's
discrete torsion. For example an $\text{O3}^+$ plane will give a $C$-type
algebra on a stack of $[0,1]$ 7-branes. With this information we can easily
infer that when a string junction of charge $[p,q]$  crosses an orientifold
3-plane of discrete torsion $(a,b)$ worldsheet parity will produce a sign
$(-1)^{a p -  b q}$ on the string state. In the following we will generalize
this to other S-folds. As a final comment, we note that when mutually
non-local 7-branes are present, we find that all that matters is whether
discrete torsion is switched on or not; this is different from the situation
with all 7-branes mutually local.  In particular, when all 7-branes are
mutually local then the spectrum is ``blind'' to some sector of discrete
torsion; for example, when all 7-branes are mutually local D7s then the
Ramond--Ramond component of the discrete torsion cannot be detected by the
7-branes.

\subsection{S-fold Projection}

\begin{figure}[t!]
\begin{center}
\includegraphics[scale=.8]{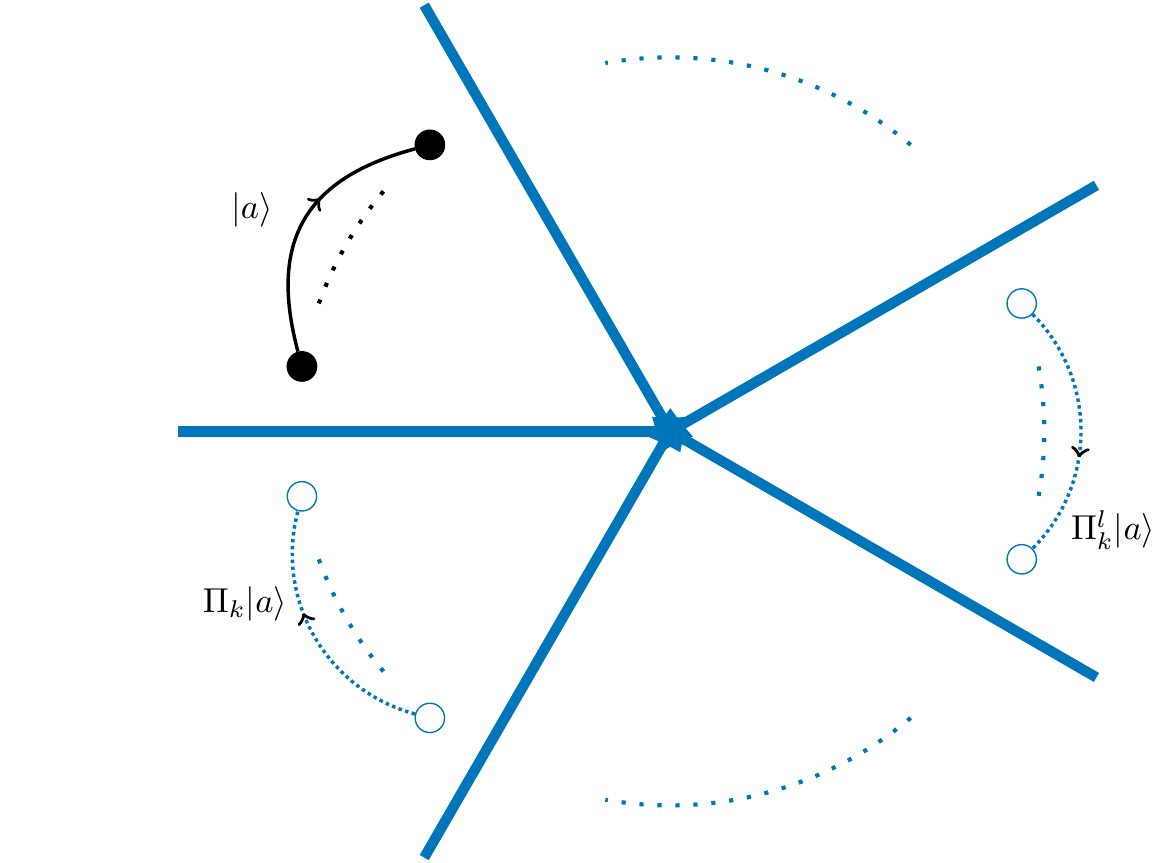}
\end{center}
\caption{Projection rules for S-fold planes acting on string junctions. We denote the orientifold image branes by
open shapes, and image strings by dashed blue lines.}
\label{fig:Sfold}
\end{figure}

In the following we will consider different $\mathbb Z_k$ projections on the
set of string junctions. To get invariant states we will call $\Pi_k$ the generator
of the $\mathbb Z_k$ action on the string state and we will sum over the $\mathbb
Z_k$ images to get the states after projection, meaning that we shall consider
the combination
\al{ \frac{1}{k} \left( \mathbb I + \sum_{l=1}^{k-1} \Pi_k^{l}\right)\,.
}
This action is considered over the generators of the complexified Lie algebra,
not on the root vectors. In particular, Lie algebra generators that are mapped
to themselves may be projected out due to some phases in $\Pi_k$. Indeed since
the only requirement for $\Pi_k$ is that its $k$-th power is the identity it
is possible to twist it by some $\mathbb Z_k$ phases corresponding to
different choices of discrete torsion; these choices were reviewed in section
\ref{sec:tor}. What needs to be fixed is the phase that is acquired by the
various junctions in the presence of discrete torsion. Note that this
information is relevant only for junctions whose root vectors are invariant
under the S-fold projection as the addition of these  phases may project them
out. We will write down the phase for a $[p,q]$-string crossing the S-fold
with torsion $(a,b)$. The phase is fixed by requiring invariance under the
torsion equivalence relations described in section \ref{sec:tor}. The various
cases are
\al{ &k= 2 \,, \qquad (-1)^{a p - b q}\,,\\
&k=3 \,, \qquad e^{\frac{2 \pi i }{3} (a p - b p -a q +b q)}\,,\\
&k=4 \,, \qquad (-1)^{    a p - b p -a q +b q}\,,
}
where we omit the case $k=6$ since no discrete torsion is available for this value.
See figure \ref{fig:Sfold} for a depiction of S-fold projection on string junction states.

In the above discussion, we have made reference to a specific duality frame.
Given that we are working at strong coupling, it is natural to ask about the
behavior of our S-fold projection under $SL(2,\mathbb{Z})$ duality
transformations. Note that while
the expression for the phase is invariant under global $SL(2,\mathbb Z)$
transformations for $k=2$, for $k > 2$ it is necessary to conjugate the
pairing between junction charges and discrete torsion under global
$SL(2,\mathbb Z)$ transformations in order to ensure that the phase is
unchanged\footnote{In practice we will conjugate the pairing for all values of $k$.}. This
should not come as a surprise as for $k>2$ we are implicitly referring to a
specific choice of an $SL(2,\mathbb Z)$ frame when discussing the torsional
fluxes: indeed the equivalence relations among discrete torsion discussed in
section \ref{sec:tor} refers to a matrix $\rho$ that is not invariant under
global $SL(2,\mathbb Z)$  transformations. Given that the product appearing in
the phase is fixed by requiring compatibility with these equivalence relations
it will necessarily be different when going to a new $SL(2,\mathbb Z)$ frame
in order to ensure that the new equivalence relations are respected.

To proceed further, we now examine the different choices of S-fold projections on different stacks of 7-branes.

\subsection{$\mathbb Z_2$ Quotients of  $E_6$}

We now turn to an analysis of $\mathbb{Z}_2$ quotients of $E_6$, namely we consider
the action of O3-planes on string junctions attached to an $E_6$ 7-brane.
We start by writing $E_6$ in a $\mathbb Z_2$-symmetric fashion.
The usual brane configuration $A^6 X C$ \cite{DeWolfe:1998zf} can be permuted
to a configuration $ A^3  C  A^3  C$. We discuss the permutations in Appendix \ref{app:BRANE}.
The set of $72$ junctions giving the roots of $E_6$ is
\begin{align} &\pm \left( {a}_i -  a_j\right)\,,\quad  1\leq j < i \leq 6\,,\nonumber\\
&\pm \left(\sum_{i=1}^3   a_i - \sum_{j=4}^6    a_j  -   a_k +  a_l +  c_1 -   c_2\right)\,,\quad  1\leq k \leq 3\,, \, 4 \leq l \leq 6\,,\nonumber\\
& \pm \left(   a_i -   a_j +  c_1 -   c_2\right) \,, \quad 1\leq k \leq 3\,, \, 4 \leq l \leq 6\,,\nonumber\\
&\pm \left(\sum_{i=1}^3   a_i - \sum_{j=4}^6    a_j +  c_1 -    c_2\right)\,,\nonumber \\
&\pm \left(\sum_{i=1}^3   a_i - \sum_{j=4}^6    a_j +2  c_1 - 2   c_2\right)\,,\nonumber \\
&\pm\left(  c_1 -   c_2\right)\,.
\label{eq:E6roots}\end{align}
A set of simple roots is given by
\al{\left\{ a_1-  a_2,  a_2-  a_3,  a_3 -   a_4,  a_4-  a_5, a_5-  a_6,  c_1-  c_2\right\}.
}
We will now turn to studying the effects of the S-fold projection, both
without and with discrete torsion (for all possible choices) turned on.
\begin{figure}[t!]
\begin{center}
\includegraphics[scale=.8]{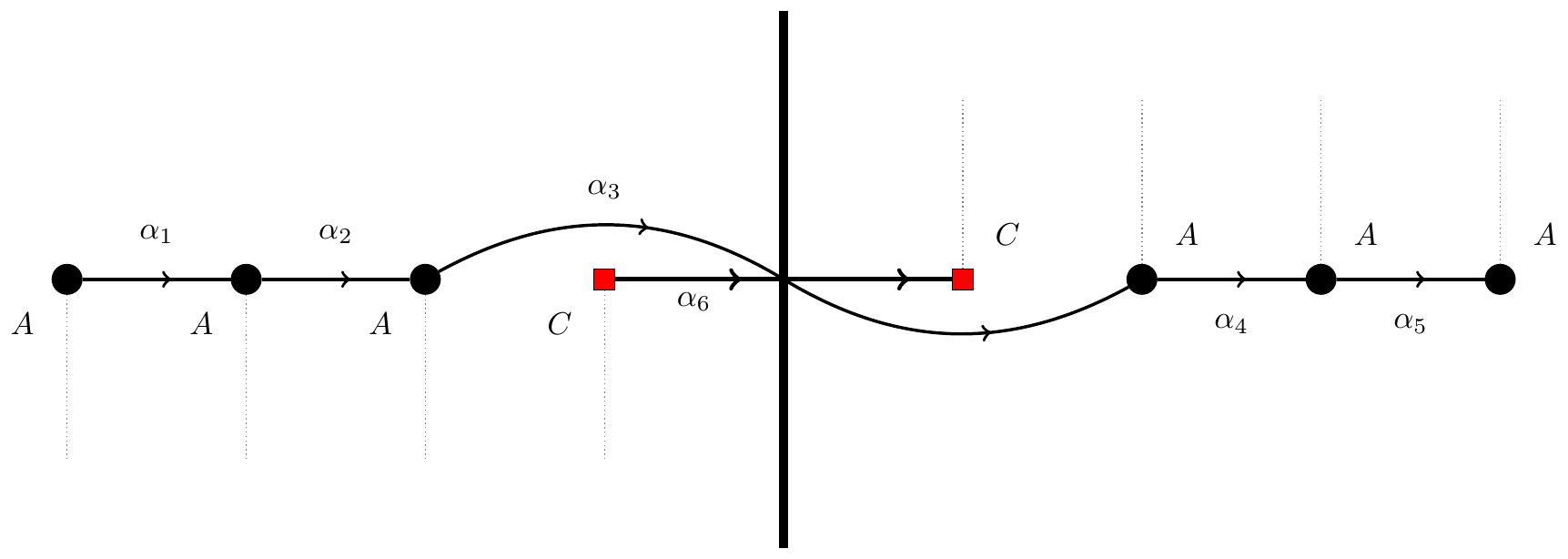}
\end{center}
\caption{$\mathbb Z_2$ symmetric configuration for $E_6$ theory.}
\label{fig:E6Z2}
\end{figure}

\subsubsection{$\mathbb Z_2$ Quotient without Discrete Torsion}

Consider first the case without any discrete torsion. After the projection
$48$ string junctions survive (see Appendix \ref{app:E6SJ} for a fully worked
example of which string junctions are projected out for the quotients of
$E_6$).  Given the symmetry of the system we can write all junctions
specifying only the charges on half the set of branes for sake of convenience.
The remaining junctions after projection are
\al{ \label{eqn:dora} &\pm \frac{1}{2}\left( {a}_i -  a_j\right)\,,\quad  1\leq j < i \leq 3\,,\nonumber\\
&\pm \frac{1}{2}\left( {a}_i +  a_j\right)\,,\quad  1\leq j < i \leq 3\,,\nonumber\\
& \pm   a_i\,, \quad 1\leq i \leq 3\,, \nonumber\\
& \pm \left(  a_i+  c_1\right), \quad 1\leq i \leq 3\,, \nonumber\\
&\pm \left(\sum_{i=1}^3   a_i -    a_j +  c_1 \right)\,, \quad 1\leq j \leq 3\,,\nonumber \\
&\pm\frac{1}{2} \left(\sum_{i=1}^3   a_i -    a_j +2  c_1 \right)\,, \quad 1\leq j \leq 3\,,\nonumber \\
&\pm \frac{1}{2}\left(\sum_{i=1}^3   a_i +    a_j +2  c_1 \right)\,, \quad 1\leq j \leq 3\,,\nonumber \\
& \pm   c_1\,,\nonumber\\
&\pm \left(\sum_{i=1}^3   a_i +   c_1\right)\,,\nonumber\\
&\pm \left(\sum_{i=1}^3   a_i + 2  c_1\right)\,.
}
This gives in total $48$ junctions, as expected for $F_4$. One choice of simple roots is
\begin{align}\left\{\frac{1}{2}\left(a_1 -   a_2\right), \frac{1}{2}\left(a_2 -   a_3\right), a_3,  c_1\right\}\,.\label{eq:F4simpleroots}
\end{align}
It is possible to check that using the intersection matrix of the brane system of $E_6$ one obtains the Cartan matrix of $F_4$,
thus indicating that the resulting algebra is $F_4$.

From the above considerations, we conclude that D3-branes probing this
S-folded 7-brane configuration will enjoy an $F_4$ global symmetry. At
first glance, this would appear to be at odds with reference
\cite{Shimizu:2017kzs} which demonstrated that for 4D $\mathcal{N} = 2$ SCFTs
with Higgs branch given by the single instanton moduli space of $F_4$ gauge
theory, there is a global inconsistency in the anomalies of the associated
theory. An important point to emphasize here, however, is that the same class
of assumptions also allows one to extract the values of various
anomalies including $\kappa_F = 5$, $a = 4/3$ and $c = 5/3$, which is rather
different from the values of references
\cite{Argyres:2015ffa,Argyres:2015gha,Argyres:2016xua,Argyres:2016xmc,Argyres:2016yzz,Martone:2020nsy,Argyres:2020wmq},
which have $\kappa_F = 6$, $a = 41/24$ and $c = 13/6$. Our analysis is compatible
with these considerations and indicate that the structure of the Higgs branch
is more subtle. Indeed, this is in line with the fact that moving the D3-brane
off the S-fold but still inside the $E_6$ 7-brane, the local spectrum of
string junction states is actually $E_6$. The brane picture indicates that it
is more appropriate, then, to view the Higgs branch moduli space for the
D3-brane as an instanton in an $E_6$ gauge theory but in the presence of a
codimension four S-fold defect.

\begin{figure}[t!]
\begin{center}
\includegraphics[scale=.8]{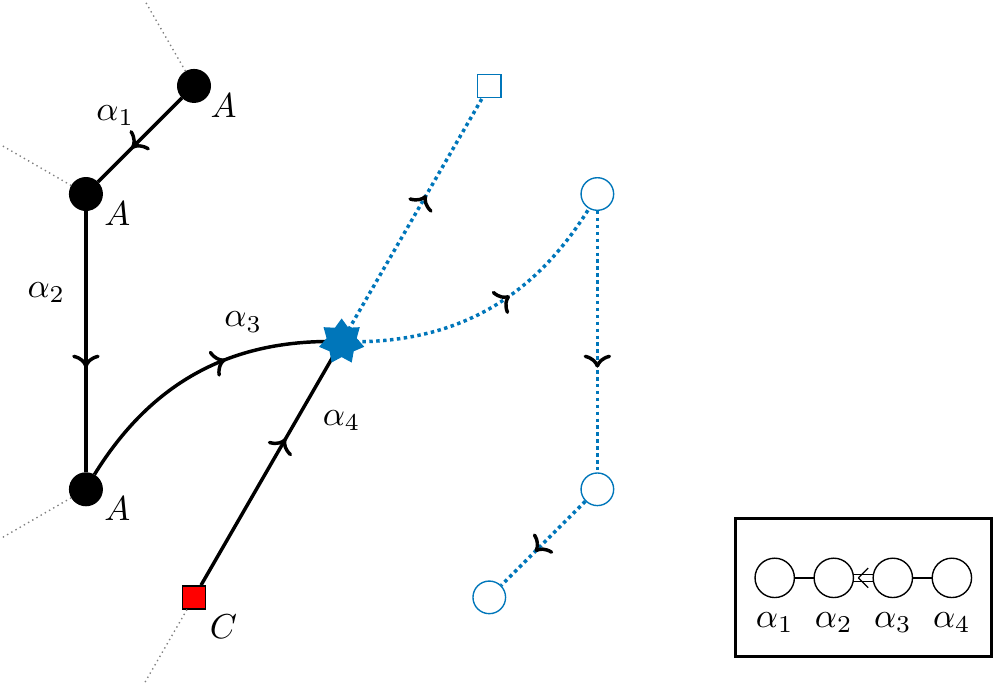}
\end{center}
\caption{String junctions for the S-fold projection of $E_6$ to $F_4$. We denote the orientifold image branes by
open shapes, and image strings by dashed blue lines.}
\label{fig:F4}
\end{figure}

\subsubsection{$\mathbb Z_2$ Quotient with Discrete Torsion}

\begin{figure}[t!]
\begin{center}
\includegraphics[scale = 1.0]{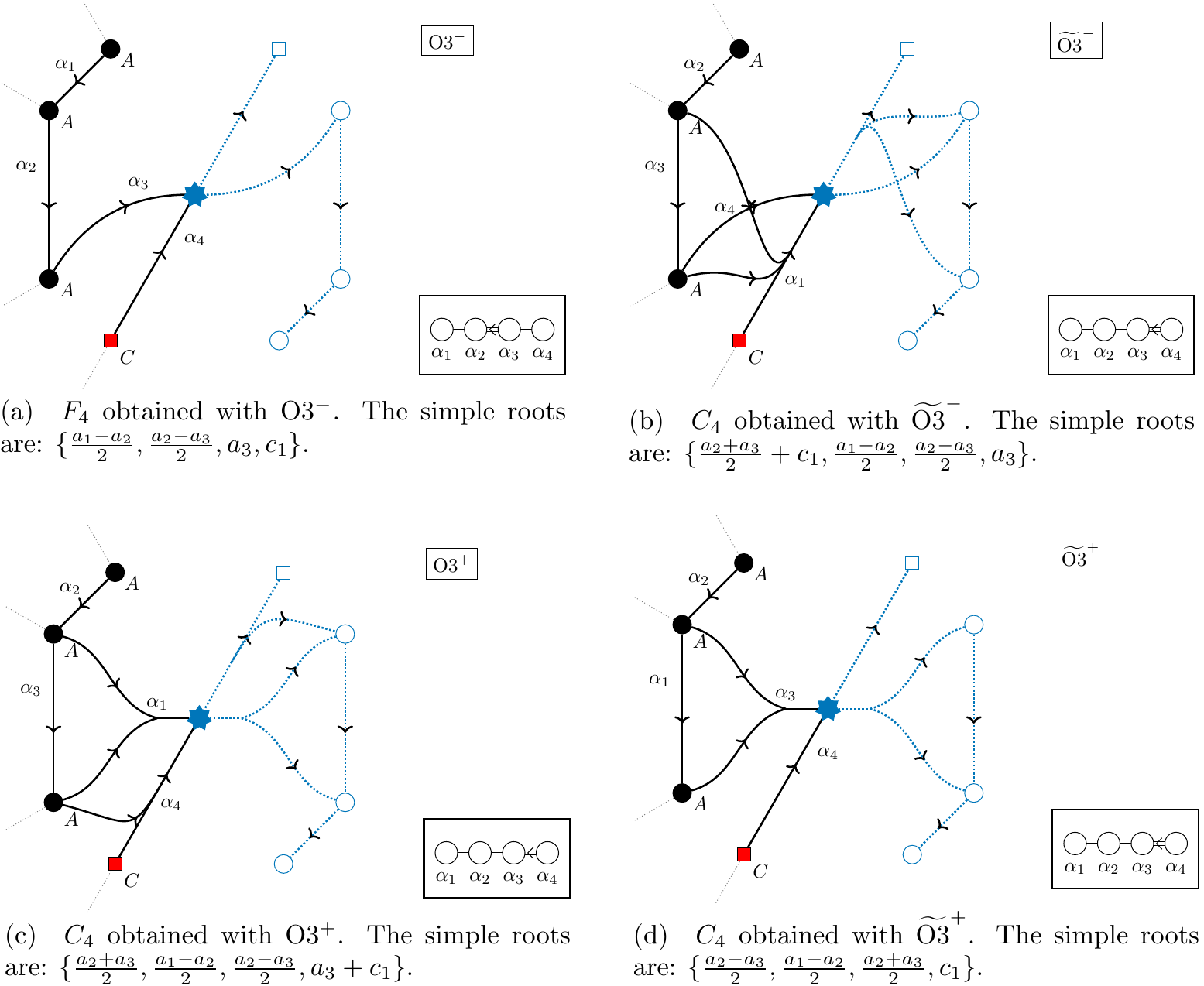}
\end{center}
\caption{Depiction of the different S-fold projections for an $E_6$ stack of 7-branes. Applying this projection results in two
physically distinct configurations, the one without discrete torsion (a), and the ones with discrete torsion (b,c,d).
We denote the orientifold image branes by open shapes, and image strings by dashed blue lines.}
\label{fig:e6sfolds}
\end{figure}

Consider next the case of an orientifold projection with discrete torsion for
string junctions attached to an $E_6$ 7-brane. We find that in all cases the
junctions that are not invariant under the $\mathbb Z_2$ action are not
affected by the torsion. These are the ones with $1/2$ factors in the formulas
written in (\ref{eqn:dora}). For all choices of non-trivial discrete torsion
we find that $16$ additional junctions are projected out, though which ones in
particular depends on the choice of the discrete torsion. This gives in all
cases a set of $32$ junctions after projection. We illustrate the different
string junction configurations which survive the S-fold projection in figure
\ref{fig:e6sfolds}. This shows that although different string junctions
survive for each choice of discrete torsion, the actual flavor symmetry
algebra realized in all these cases is the same. Moreover, this analysis
establishes that in all cases the root system is the one of $C_4$.

\subsection{$\mathbb Z_2$ Quotients of  $D_4$}

Consider next $\mathbb{Z}_2$ quotients of $D_4$. Recall that in F-theory, this is associated with a type $I_{0}^{\ast}$ fiber.
In this case, it is helpful to use the fact that the $E_6$ stack can be written as $AAA C AAAC$, so removing an A-brane from each grouping, we arrive at $AAC AAC$, the $\mathbb{Z}_2$ symmetric grouping for $D_4$. Therefore, one obtains the roots of $D_4$ by selecting the $E_6$ junctions without $  a_3$ and $  a_6$.  Moreover, for notational simplicity we shall rename $  a_{i+3}$ as $  a_{i+2}$ for $i = 1,2$. There are $24$ remaining junctions (as expected), and we list them here:
\al{ &\pm \left( {a}_i -  a_j\right)\,,\quad  1\leq j < i \leq 4\,,\nonumber\\
&\pm \left(  a_1 +   a_2 -   a_3 -  a_4 +  c_1 -   c_2\right)\,,\nonumber\\
& \pm \left(   a_i -   a_j +  c_1 -   c_2\right) \,, \quad 1\leq i \leq 2\,, \, 3 \leq j \leq 4\,,\nonumber\\
&\pm\left(  c_1 -   c_2\right)\,.
}
A set of simple roots is given by
\al{\left\{a_1-  a_2, a_2 -   a_3,a_3-  a_4,  c_1-  c_2\right\}.
}
Let us now turn to the different S-fold (really orientifold)
projections in this case.

\subsubsection{$\mathbb Z_2$ Quotient without Discrete Torsion}

Consider first the S-fold projection of $D_4$ without discrete torsion. In this case we
find $18$ junctions which survive, and this is the dimension of the root system of both $B_3$ and $C_3$.
As before, after the quotient we can write the junction specifying the charges on only half the set of branes.
The junctions after the projection are
\al{ &\pm \frac{1}{2}\left(  a_1 -   a_2\right)\,,\nonumber\\
&\pm\frac{1}{2} \left( {a}_1 +  a_2\right)\,,\nonumber\\
& \pm   a_i\,, \quad 1\leq i \leq 2\,, \nonumber\\
& \pm \left(  a_i+  c_1\right), \quad 1\leq i \leq 2\,, \nonumber\\
&\pm \left(  a_1 +   a_2 +  c_1 \right)\,,\nonumber \\
&\pm \frac{1}{2}\left(  a_1 +   a_2 +2  c_1 \right)\,,\nonumber \\
& \pm   c_1\,.
}
Computing the Cartan matrix we finds it corresponds to the Lie algebra $B_3$. One choice of simple roots is
\al{\left\{\frac{1}{2}\left(a_1 -   a_2\right), a_2, c_1\right\}\,.
}
As an additional comment, we observe that the above brane construction can be
viewed as specifying a mass deformation from a theory with $F_4$ global
symmetry to one with $B_3$ symmetry. This is indeed precisely the sort of
deformation observed from purely bottom up considerations in reference
\cite{Argyres:2015gha}. One can see this mass deformation as a blue arrow
between the $[II^*, F_4]$ and the $[III^*, B_3]$ theories in figure \ref{fig:N2rankone}.

\subsubsection{$\mathbb Z_2$ Quotient with Discrete Torsion}

Consider next the case of $D_4$ 7-branes in the presence of an orientifold (i.e. $\mathbb{Z}_2$ S-fold)
with discrete torsion. The result is that after the projection, $10$ string junctions survive for all different choices of
discrete torsion other than the trivial one. In all cases the resulting algebra is $C_2 \oplus A_1$.

\subsection{$\mathbb Z_2$ Quotients of  $H_2$}

The final case allowed with the $\mathbb Z_2$ S-fold is an $H_2$ 7-brane, namely a type $IV$ fiber at the origin.
We can obtain it by starting from the $AAC AAC$ realization of the $D_4$ case and dropping an $A$-brane from
both stacks, resulting in the configuration $ACAC$. The junctions can thus be obtained from the $D_4$ ones and this yields:
\al{ &\pm \left( {a}_1 -  a_2\right)\,,\nonumber\\
& \pm \left(   a_1 -   a_2 +  c_1 -   c_2\right) \,,\nonumber\\
&\pm\left(  c_1 -   c_2\right)\,.
}
A set of simple roots is given by:
\al{\{a_1-  a_2, c_1 -   c_2\}\,.
}
So, we get $6$ junctions as expected for $H_2$, giving an $A_2$ algebra.
Let us now turn to S-fold projections of this flavor symmetry.

\subsubsection{$\mathbb Z_2$ Quotient without Discrete Torsion}

Consider first the $\mathbb{Z}_2$ quotient without discrete torsion of an
$H_2$ flavor 7-brane. In this case, it is interesting to note that all string
junctions are invariant under the $\mathbb Z_2$ action when there is no
discrete torsion.  Consequently, we retain the same flavor symmetry algebra.
In the context of 4D $\mathcal{N} = 2$ SCFTs \cite{Argyres:2015gha}, we
observe that we can also consider the associated flow, via mass deformation,
from $[III^*,B_3]$ to $[IV^*,A_2]$, as in figure \ref{fig:N2rankone}, which is
compatible with our brane picture.

\subsubsection{$\mathbb Z_2$ Quotients with Discrete Torsion}

We next consider the $\mathbb Z_2$ projection with discrete torsion of the $H_2$ theory.
The result is that after the projection, $2$ string junctions survive for all different
choices of discrete torsion other than the trivial one. In all cases the resulting algebra
is $A_1 \oplus U(1)$. Here we observe the appearance of a $U(1)$ factor in the symmetry algebra. We see this
since there are string junctions stretched to just the $C$ brane of the configuration $A^{3} C$ realizing $H_2$ and its
subsequent $\mathbb{Z}_2$ quotient. This is also in accord with the quotient group action on the symmetry algebra
of the parent theory.

\subsection{$\mathbb Z_3$ Quotients of $D_4$}

As we already saw in section \ref{sec:7BRANE}, the $D_4$ configuration of 7-branes also admits a $\mathbb{Z}_3$ S-fold quotient. Here,
we study the resulting algebras both in the absence and in the presence of discrete torsion. To proceed, we observe that the $\mathbb Z_3$ symmetric choice of branes is $AABBDD$ where $D$ is a $[0,1]$-brane. In this presentation the junctions giving the root system of $D_4$ are
\al{ &\pm\left(a_1 -a_2\right)\,,\quad \pm\left(b_1 -b_2\right)\,, \quad \pm\left(d_1 -d_2\right)\,,\nonumber\\
&\pm \left(a_i - b_j -d_k\right)\,, 	\quad i=1,2\,, \ j = 1,2\,,\ k = 1,2\,,\quad \nonumber\\
&\pm\left(a_1 + a_2 -b_1-b_2-d_1-d_2\right)\,.
}
One choice of simple roots is
\al{\{-a_1+a_2,a_1-b_1-d_1,d_1-d_2,b_1-b_2\}\,.
}

\subsubsection{$\mathbb Z_3$ Quotient without Discrete Torsion}

With this in place, we are ready to discuss $\mathbb{Z}_3$ S-fold projections of $D_4$ 7-branes. Consider first the case of
S-fold projections without discrete torsion. The $\mathbb Z_3$ action maps the branes as follows
\al{a_i \rightarrow -b_i\,, \quad b_i \rightarrow d_i \,, \quad d_i \rightarrow -a_i\,.
}
After the projection the remaining junctions are
\al{&\pm\left(a_i - b_i -d_i\right)\,, \quad i = 1,2\,,\nonumber\\
& \pm\frac{1}{3}\left(-a_1 + a_2 +b_1-b_2+d_1-d_2\right)\,,\quad  \pm\left(a_1 + a_2 -b_1-b_2-d_1-d_2\right)\,,\nonumber\\
&\pm \frac{1}{3}\left( 2a_1 +a_2 -2 b_1 - b_2 -2 d_1 - d_2\right)\,, \quad \pm\frac{1}{3}\left( a_1 +2a_2 - b_1 -2 b_2 - d_1 -2 d_2\right)\,.
}
The simple roots after projection can be chosen to be
\al{\left\{\frac{1}{3}\left(-a_1 + a_2 +b_1-b_2+d_1-d_2\right),a_1 - b_1 -d_1\right\}\,,
}
whose intersection gives the Cartan matrix of $G_2$, which matches to the $[II^*,G_2]$ theory of reference
\cite{Argyres:2015gha}.

\subsubsection{$\mathbb Z_3$ Quotients with Discrete Torsion}

We next consider the $\mathbb Z_3$ projection with discrete torsion of the $D_4$ theory. In this case the reason why some junctions may be projected out is that after summing over the $\Pi_3$ images they get a factor $1 + \zeta + \zeta^2 = 0$ where $\zeta$ is a primitive third
root of unity. One can check that for both choices of discrete torsion the junctions $\pm (a_i -b_i-d_i)$ for $i=1,2$ and $\pm (a_1+a_2-b_1-b_2-d_1-d_2)$ are projected out. This leaves in total 6 junctions giving the $A_2$ algebra.\footnote{Going from $G_2$ to $A_2$ follows because the root system of $G_2$ is nothing but the root system of $A_2$ with the addition of the weights of the $\mathbf 3$ and $\bar{\mathbf 3}$ representations. Including discrete torsion projects out these vectors leaving only $A_2$ behind.}

\subsection{$\mathbb Z_3$ Quotients of $H_1$}

Let us now turn to $\mathbb{Z}_3$ quotients of the $H_1$ stack of 7-branes. We can use our analysis of the $D_4$ stack of 7-branes to
aid in this analysis. To this end, we begin with the realization of the $D_4$ algebra using the $\mathbb{Z}_3$ symmetric
stack $AABBDD$. We get to the $H_1$ stack by removing one $A$ brane, one $B$ brane and one $D$ brane. The remaining junctions are
\al{\pm (a - b - d)\,,
}
thus giving an $A_1$ algebra.

\subsubsection{$\mathbb Z_3$ Quotient without Discrete Torsion}

Consider first the $\mathbb{Z}_3$ S-fold projection in the absence of discrete
torsion. This junction is already invariant under the $\mathbb Z_3$ quotient
suggesting that the theory can be identified with the $[III^*,A_1]$ of
\cite{Argyres:2015gha}. Indeed there is a flow $[II^*,G_2] \rightarrow [III^*,A_1]$ for
the corresponding 4D $\mathcal{N} = 2$ SCFTs.

\subsubsection{$\mathbb Z_3$ Quotients with Discrete Torsion}

Next consider the $\mathbb{Z}_3$ S-fold projection with discrete torsion. In
both cases of $\mathbb Z_3$ discrete torsion there are no junctions surviving
leaving only one single Cartan generator behind. The flavor symmetry is
therefore simply $U(1)$.

\subsection{$\mathbb Z_4$ Quotients of $H_2$}

We next turn to the $\mathbb{Z}_4$ S-fold projection of the $H_2$ stack of 7-branes.
The brane system can be conjugated to a $DADA$ system  where again $D$ is a $[0,1]$-brane. The junctions giving the roots are
\al{ \pm\left(a_1-a_2\right)\,, \quad \pm\left(d_1-d_2\right)\,, \quad \pm\left(a_1-a_2 + d_1-d_2\right)\,.
}
\subsubsection{$\mathbb Z_4$ Quotient without Discrete Torsion}

Consider first the $\mathbb{Z}_4$ S-fold projection without discrete torsion on the $H_2$ stack of 7-branes.
The $\mathbb Z_4$ projection maps
\al{ a_1 \rightarrow d_1\,, \quad d_1 \rightarrow- a_2 \,, \quad a_2 \rightarrow d_2\,, \quad  d_2 \rightarrow -a_1\,.
}
After projection one finds only the junctions
\al{ \pm \left(a_1 + d_1 -a_2-d_2 \right)\,.
}
The algebra is therefore $A_1$ thus giving the $[II^*,B_1]$
theory.\footnote{Note that at the level of Lie algebras we have $A_1 \simeq
B_1$.}

\subsubsection{$\mathbb Z_4$ Quotient with Discrete Torsion}

In the case of the $\mathbb{Z}_4$ S-fold projection with discrete torsion of
the $H_2$ stack of 7-branes, we find by a similar analysis that the algebra is $A_1$, i.e. there is no distinction
in the flavor symmetry algebras for the cases with and without discrete
torsion.

\subsection{Collection of Flavor Symmetry Algebras}

In this section we collect our results on the resulting flavor symmetry
algebras. First, we remind the reader that the particular non-zero values of the discrete torsion are
irrelevant; the spectrum of physical states, as determined from the string junctions, is
identical for all cases with non-zero discrete torsion. We then
summarize the different algebras and a choice of root system in tables
\ref{tab:Z2}, \ref{tab:Z3}, \ref{tab:Z4}. In table \ref{tab:ENHANCO} we summarize
the relevant patterns, indicating quotients without discrete torsion as $\mathbb{Z}_k$
and those with discrete torsion as $\widehat{\mathbb{Z}}_k$.

\renewcommand{\arraystretch}{1.5}

\begin{table}[]
    \centering
    \begin{tabular}{|c|c|c|c|}\hline
        S-fold & $E_6/\bZ_2$ & $D_4/\bZ_2$ & $H_2/\bZ_2$  \\ \hline\hline
        $\mathrm{O3}^-$  & $F_4: $ & $B_3:$ & $A_2:$\\
         & $\{\frac{a_1-a_2}{2}, \frac{a_2-a_3}{2}, a_3, c_1 \}$ &$ \{\frac{a_1-a_2}{2}, a_2, c_1 \}$  &  $\{a_1 , c_1 \}$
        \\ \hline\hline
        $\mathrm{\widetilde{O3}}^-$  &
        $C_4:$ & $C_2 \oplus A_1:$ & $A_1 \oplus U(1):$\\
        & $\{\frac{a_2+a_3}{2}+c_1, \frac{a_1-a_2}{2}, \frac{a_2-a_3}{2}, a_3 \}$ & $ \{ \frac{a_1 - a_2}{2}, a_2 \}\oplus \{\frac{a_1 + a_2}{2}+c_1 \}$ & $ \{a_1 \}$
        \\ \hline
        $\mathrm{O3}^+$ &
        $C_4:$ & $C_2 \oplus A_1: $ & $A_1 \oplus U(1): $ \\
        & $ \{\frac{a_2+a_3}{2}, \frac{a_1-a_2}{2}, \frac{a_2-a_3}{2}, a_3+c_1 \}$ & $\{ \frac{a_1 - a_2}{2}, c_1 + a_2 \} \oplus \{\frac{a_1 + a_2}{2} \} $ &  $\{a_1 + c_1 \}$
        \\ \hline
        $\mathrm{\widetilde{O3}}^+$  & $C_4:$ & $C_2 \oplus A_1:$ & $A_1 \oplus U(1):$ \\
         & $ \{\frac{a_2-a_3}{2}, \frac{a_1-a_2}{2}, \frac{a_2+a_3}{2}, c_1 \}$ & $ \{ \frac{a_1 - a_2}{2}, -a_1-a_2-c_1 \}\oplus \{\frac{a_1 - a_2}{2} \}$ &$ \{c_1 \}$
        \\ \hline
    \end{tabular}
    \caption{Simple roots of $\mathbb{Z}_2$ S-folds (i.e. orientifold projection) with all possible choices of discrete torsion.}
    \label{tab:Z2}
\end{table}

\begin{table}[]
    \centering
    \begin{tabular}{|c|c|c|}\hline
        S-fold & $D_4/\bZ_3$ & $H_1/\bZ_3$  \\ \hline\hline
    Trivial & $G_2:$ & $A_1:$\\
    torsion& $\{\frac{1}{3}\left(-a_{1}+a_{2}+b_{1}-b_{2}+d_{1}-d_{2}\right), a_{1}-b_{1}-d_{1}\}$ &  $\{a - b - d\}$
    \\\hline\hline
    Non-trivial & $A_2:$ & $U(1)$ \\
    torsion& $\{\frac{1}{3}\left(-a_{1}+a_{2}+b_{1}-b_{2}+d_{1}-d_{2}\right),\frac{1}{3}\left(2 a_{1}+a_{2}-2 b_{1}-b_{2}-2 d_{1}-d_{2}\right)\}$ &
    \\\hline
    \end{tabular}
    \caption{Simple roots of $\mathbb{Z}_3$ S-folds with all possible choices of discrete torsion.}
    \label{tab:Z3}
\end{table}

\begin{table}[]
    \centering
    \begin{tabular}{|c|c|c|}\hline
        S-fold & $H_2/\bZ_4$\\ \hline\hline
    Trivial torsion & $A_1: \{a_1 + d_1 - a_2 - d_2\}$
    \\\hline
    Non-trivial torsion & $A_1: \{a_1 + d_1 - a_2 - d_2\}$
    \\\hline
    \end{tabular}
    \caption{Simple roots of $\mathbb{Z}_4$ S-folds with all possible choices of discrete torsion. Here having non-trivial torsion does not affect the gauge algebra or the simple root system.}
    \label{tab:Z4}
\end{table}

\renewcommand{\arraystretch}{1}

\begin{table}[]
\centering
\begin{tabular}{c c c c c c c}
\textbf{parent } & $\mathbb Z_2$ & $\widehat{\mathbb Z}_2$ & $\mathbb Z_3$ & $\widehat{\mathbb Z}_3$ & $\mathbb Z_4$ & $\widehat{\mathbb Z}_4$\\
\hline
$E_8$ \\
$E_7$\\
$E_6$ & $F_4$ & $C_4 $\\
$D_4$ & $B_3$ & $ C_2 \oplus A_1 $ & $G_2$ & $A_2$\\
$H_2$ & $A_2$ & $A_1 \oplus U(1) $ & & & $A_1$ & $A_1$\\
$H_1$ & & & $A_1$ & $U_1 $\\
$H_0$
\end{tabular}
    \caption{Summary of symmetry algebras obtained from an S-fold projection of a parent stack of 7-branes. We find that there are two qualitative quotients, based on $\mathbb{Z}_k$ without discrete torsion, and based on $\widehat{Z}_{k}$ with discrete torsion.}
    \label{tab:ENHANCO}
\end{table}

The aforementioned flavor algebras are always realized on the worldvolume of
the 7-branes and for all ranks of the SCFT. However it is expected that in the case of
rank one theories, a quotient with discrete torsion can result in an enhancement of the geometric
flavor symmetry and that realized by a 7-brane. This geometric symmetry is $SU(2)$
for $\widehat{\mathbb{Z}}_2$ quotients and $U(1)$ for the other $\widehat{\mathbb{Z}}_{k}$ quotients.
We can determine that there is likely an enhancement when the level of the
$SU(2)$ and the level of the 7-brane flavor symmetry (both
of which we can calculate) match. The expected enhancements \cite{Argyres:2016yzz} are:
\begin{itemize}
\item[-] For the $\widehat {\mathbb Z}_2$ quotient of $E_6$ the rank one theory is expected to have $C_5$ flavor symmetry;
\item[-] For the $\widehat {\mathbb Z}_2$ quotient of $D_4$ the rank one theory is expected to have $C_3 \oplus A_1$ flavor symmetry;
\item[-] For the $\widehat {\mathbb Z}_2$ quotient of $H_2$ the rank one theory is expected to have $C_2 \oplus U_1$ flavor symmetry;
\item[-] For the $\widehat {\mathbb Z}_3$ quotient of $D_4$ the rank one theory is expected to have $A_3 \rtimes \mathbb Z_2$ flavor symmetry;
\item[-] For the $\widehat {\mathbb Z}_3$ quotient of $H_1$ the rank one theory is expected to have $A_1 \oplus U_1 \rtimes \mathbb Z_2$ flavor symmetry;
\item[-] For the $\widehat {\mathbb Z}_4$ quotient of $H_2$ the rank one theory is expected to have $A_2 \rtimes \mathbb Z_2$ flavor symmetry.
\end{itemize}

\subsection{Admissible Representations}

So far we have focused on the structure of the Lie algebra of the flavor symmetry.
The string junction picture also allows us to access the admissible representations.
We will discuss only the cases where the center of the simply connected group of a given Lie algebra is non-trivial. We begin by first discussing S-fold projections without discrete torsion, and then turn to the case of examples with discrete torsion.
If there happen to be other sources of flavor symmetries, this can lead to additional global structure. For example, $E_8$ has an $E_6 \times SU(3) / \mathbb{Z}_3$ subgroup, but also has representations in the $(\mathbf{27} , \mathbf{3})$. If we ignore the $SU(3)$ factor, then we would loosely refer to this as realizing an $E_6$ group. In the probe D3-brane theories,
we also know that there is an $SU(2)$ flavor symmetry associated with symmetries internal to the 7-brane but transverse to the D3-brane,
so determining the full structure of the 4D flavor symmetry must reference this feature as well. We leave this determination for future work. What we can assert from the string junction picture is whether we see evidence for a given type of representation, and so to indicate this information we will mildly abuse terminology and refer to $G_{\mathrm{rep}}$ as specifying the ``the flavor group'' and its admissible representations.

\subsubsection{S-fold Projections without Discrete Torsion}

We now turn to S-fold projections without discrete torsion in which, for a given Lie algebra, the associated
simply connected Lie group has a non-trivial center. This limits us to the following cases:

\begin{itemize}
\item[-] The $\mathbb Z_2$ quotient of a $D_4$ stack of 7-branes without discrete torsion yields a $B_3$ algebra, which means that the flavor group is either $Spin(7)$ or $Spin(7)/\mathbb Z_2 \simeq SO(7)$. One quick way to check which representations are allowed is to use the fact that the $B_3$ theory descends from the $F_4$ theory. Decomposing the adjoint of $F_4$ one finds
\al{ F_4 &\rightarrow Spin(7) \otimes SO(2)\\
\mathbf {52} & \rightarrow \mathbf 1_0 \oplus \mathbf 7_{2} \oplus \mathbf 7_{-2} \oplus \mathbf {21}_0 \oplus \mathbf 8_1 \oplus \mathbf 8_{-1}\,.
}
Note that the $\mathbf 8$ is the spinor representation of $Spin(7)$ so indeed the flavor group is $Spin(7)$.
\item[-] The $\mathbb Z_2$ quotient of a $H_2$ stack of 7-branes without discrete torsion yields an $A_2$ algebra, which means that the flavor group is either $SU(3)$ or $SU(3)/\mathbb Z_3 \simeq PSU(3)$. Similarly to the previous case we can use the fact that the $A_2$ theory descends from the $B_3$ theory. Decomposing the adjoint of $Spin(7)$ one finds
\al{ Spin(7) &\rightarrow SU(3) \otimes U(1)\\
\mathbf {21} & \rightarrow \mathbf {1}_0 \oplus \mathbf 8_0\oplus \mathbf 3_4 \oplus \mathbf{\bar 3}_{-4}\oplus \mathbf 3_2 \oplus \mathbf{\bar 3}_{-2}\,.
}
Since the $\mathbf 3$ representation of $A_2$ is present this fixes the flavor symmetry group to be $SU(3)$.

\item[-] The $\mathbb Z_3$ quotient of an $H_1$ theory without discrete torsion gives the flavor algebra $A_1$, which means that the flavor group could be either $SU(2)$ or $SU(2)/\mathbb Z_2 \simeq SO(3)$. We can follow the logic outlined before noting that this theory comes from the $G_2$ theory. Decomposing  the adjoint of $G_2$ we find
\al{ G_2 & \rightarrow SU(2) \otimes SU(2)\,,\\
\mathbf{14} & \rightarrow (\mathbf 3,\mathbf 1) \oplus (\mathbf 1, \mathbf 3) \oplus (\mathbf 4,\mathbf 2)\,.
}
It is possible to check by computing the charges of the junctions that after breaking $G_2$ the junctions lie in the $\mathbf 4$ representation of the unbroken group, implying that this group is $SU(2)$ rather than $SO(3)$ given that the $\mathbf 4$ is charged under the center.
\end{itemize}

\subsubsection{S-fold Projections with Discrete Torsion}

Let us now turn to the related case of S-fold projections with discrete torsion. Again, we confine our analysis to
those Lie algebras which have a simply connected Lie group with non-trivial center. The relevant cases are:

\begin{itemize}
\item[-] The $\mathbb Z_2$ quotient of the $E_6$ theory with discrete torsion gives the flavor algebra $C_4$, which means that the flavor group can be either $USp(8)$ or $USp(8)/\mathbb Z_2$. In this case we note that all junctions must descend from junctions of the parent $E_6$ theory and its weight lattice is generated by the junctions giving the $\mathbf {27}$ representation. Decomposing it we find
\al{ E_6 & \rightarrow USp(8)\,,\\
\mathbf{27} & \rightarrow \mathbf{27}\,.
}
The $\mathbf {27}$ of $USp(8)$ is the two-index anti-symmetric representation which is not charged under the center. This implies that no junctions charged under the center can be generated, implying that the flavor group is $USp(8)/\mathbb Z_2$.

\item[-] The $\mathbb Z_2$ quotient of the $D_4$ theory with discrete torsion gives the flavor algebra $C_2 \oplus A_1$. Here there are various possibilities for the global structure of the gauge group. Knowing that this theory descends from the $C_4$ theory we can decompose the adjoint of $C_4$
\al{ USp(8) & \rightarrow USp(4) \otimes SU(2) \otimes U(1)\,,\\
\mathbf {36} & \rightarrow (\mathbf 4,\mathbf 2)_1 \oplus (\mathbf 4 , \mathbf 2)_{-1} \oplus (\mathbf 1,\mathbf 3)_0 \oplus (\mathbf 1,\mathbf 3)_{2} \oplus (\mathbf 1,\mathbf 3)_{-2} \oplus (\mathbf 1, \mathbf 1)_0 \oplus (\mathbf{10},\mathbf 1)_0\nonumber\,.
}
We see that the only representations charged under the center of $USp(4)$ and $SU(2)$ appear together, which suggests that the group is $\left(USp(4) \otimes SU(2)\right)/\mathbb Z_2$. Note that other quotients like for instance $USp(4)/\mathbb Z_2 \otimes SU(2)/\mathbb Z_2$ are not compatible with the representations appearing given that the fundamental representations of $USp(4)$ and $SU(2)$ appear in the previous decomposition. Following a similar logic starting from the $\mathbf {27}$ representation of $USp(8)$ which is the smallest representation available confirms this result.

\item[-] The $\mathbb Z_2$ quotient of the $D_4$ theory with discrete torsion gives the flavor algebra $A_1 \oplus U_1$. In this case we can decompose the adjoint of $C_2 \oplus A_1$ as
\al{ USp(4) \otimes SU(2) &\rightarrow SU(2) \otimes U(1)_a \otimes U(1)_b\,,\\
(\mathbf{10},\mathbf 1)\oplus (\mathbf 1, \mathbf 3) &\rightarrow  \mathbf 1_{(0,0)}\oplus \mathbf 1_{(0,0)} \oplus \mathbf 3_{(0,0)}\oplus \left(\mathbf 2_{(1,1)} \oplus   \mathbf 1_{(2,2)} \oplus \mathbf 1_{(2,-2)} \oplus \text{h.c}\right)\,.
}
The broken generator is $U(1)_b$ leaving $SU(2) \otimes U(1)_a$. Therefore the flavor symmetry group seems to be  $\left(SU(2) \otimes U(1)\right)/\mathbb Z_2$. The conclusion does not change when looking at other representations of $\left(USp(4) \otimes SU(2)\right)/\mathbb Z_2$.

\item[-] The $\mathbb Z_3$ quotient of the $D_4$ theory with discrete torsion gives the flavor algebra $A_2$, which means that the flavor group can be either $SU(3)$ or $SU(3)/\mathbb Z_3 \simeq PSU(3)$. In this case we note that all junctions must descend from junctions of the parent $E_6$ theory and its weight lattice is generated by the junctions giving the $\mathbf {8_s}$, the $\mathbf {8_c}$ and the $\mathbf {8_v}$ representations. Decomposing them we find
\al{ Spin(8) & \rightarrow SU(3)\,,\\
\mathbf{8_s} & \rightarrow \mathbf{8}\,,\\
\mathbf{8_c} & \rightarrow \mathbf{8}\,,\\
\mathbf{8_v} & \rightarrow \mathbf{8}\,.
}
The $\mathbf 8$ representation of $A_2$ is of course the adjoint which is uncharged under the center. This means that no representation charged under the center is present, giving the flavor symmetry $PSU(3)$.

\item[-] The $\mathbb Z_4$ quotient of the $H_2$ theory with discrete torsion gives the flavor algebra $A_1$, which means that the flavor group can be either $SU(2)$ or $SU(2)/\mathbb Z_2 \simeq SO(3)$. In this case we note that all junctions must descend from junctions of the parent $H_2$ theory and its weight lattice is generated by the junctions giving the $\mathbf 3$ representation. Decomposing them we find
\al{ SU(3) & \rightarrow SU(2)\,,\\
\mathbf{3} & \rightarrow \mathbf{3}\,.
}
The $\mathbf 3$ representation of $A_1$ is of course the adjoint which is uncharged under the center. This means that no representation charged under the center is present giving the flavor symmetry $SO(3)$.

\end{itemize}

\section{F-theory and S-folds with Discrete Torsion}\label{sec:DISCRETE}

One useful application of the F-theory construction is that it allows one to read
off the Seiberg--Witten curve from the geometry for the rank one
theories. However, as we stressed before, this procedure works only in the
absence of discrete torsion. Given this identification between geometry and
the low-energy field theory data it is tempting to push this identification
beyond the case without discrete torsion. We propose that the F-theory
geometry in the presence of discrete torsion is the Seiberg--Witten curve of
the theory on a single probe D3-brane. In this section we will list all the
maximally mass deformed Seiberg--Witten curves from \cite{Argyres:2015gha} for
the various theories we obtained in the presence of discrete torsion. One
subtle point is that in the case of a single D3-brane, there can be additional
enhancements in the flavor symmetry relative to the case of multiple
D3-branes. In these cases, we interpret the F-theory geometry as the one
obtained by taking a mass deformation of the enhanced symmetry algebra which
takes us to the generic flavor symmetry, and then taking a further scaling
limit so that the terms with the mass deformation are scaled out. In all
cases, this is associated with the degree two Casimir invariants of the flavor
symmetry algebra. In what follows, we leave this operation implicit in our
discussion. With notation as earlier, we use the Coulomb branch parameter $u$
to indicate the directions transverse to the 7-brane in the quotiented
geometry.

\begin{itemize}
\item[-] The Seiberg--Witten curve for the $\mathbb Z_2$ quotient with discrete torsion of the $E_6$ theory is
\al{ y^2 = x^3 &+ 3 x \left[2 u^3 M_2 +u^2 \left(M_4^2 -2 M_8\right)+2 u M_4 M_{10}-M_{10}^2\right]\nonumber\\
&+ 2 \left[u^5 + u^4 M_6+u^3 \left(2 M_4^3-3 M_4 M_8 -3 M_2 M_{10}\right)\right.\\
& \left.+3u^2 M_8 M_{10} -3 u M_4 M_{10}^2 +M_{10}^3 \right]\nonumber\,.
}

\item[-] The Seiberg--Witten curve for the $\mathbb Z_2$ quotient with
  discrete torsion of the $D_4$ theory is
\al{ y^2 = x^3 &+  x \left[ 12 u^3 -u^2\left(M_4 + 4 M_2^2\right)+12 u M_2 M_6-3 M_6^2\right]\nonumber\\
    &-12 u^4 \left(2 M_2 + 3 \widetilde{M}_2 \right)+2 u^3 \left(M_2 M_4 + 6 M_6\right)\\
& -u^2 \left(16 M_2^2 + M_4\right)M_6+12u M_2 M_6^2 -2 M_6^3\nonumber\,.
}
Note the presence of two independent degree two Casimirs, $M_2$ and
    $\widetilde{M}_2$. This occurs whenever the flavor symmetry is
    semi-simple, in this case it is $C_3 \oplus A_1$.

\item[-] The Seiberg--Witten curve for the $\mathbb Z_2$ quotient with discrete torsion of the $H_2$ theory is
\al{ y^2 = x^3 &-  x \left[ 3 u^2 \left( M_2 + M_1^2 \right)+12u M_1 M_4+3 M_4^2\right]\nonumber\\
&-864 u^4 + 2 u^3 M_1 \left(M_1^2 -3 M_2\right)-3 u^2 (5M_1^2 + M_2)M_4\\
&-12 u M_1 M_4^2 -2 M_4^3 \nonumber\,.
}

\item[-] The Seiberg--Witten curve for the $\mathbb Z_3$ quotient with discrete torsion of the $D_4$ theory is
\al{ y^2 = x^3 +3 x u^2 \left(2 u M_2 - M_4^2 \right)+2u^3\left(u^2 +M_4^3+ u M_6\right)\,.
}
This was identified in \cite{Argyres:2016xua} and reproduces the curve already found in \cite{Chacaltana:2016shw}.

\item[-] The Seiberg--Witten curve for the $\mathbb Z_3$ quotient with discrete torsion of the $H_1$ theory is
  \al{y^2 = x^3 +3 x \left(u^3 -u^2 \widetilde{M}_2^2 \right)+ 2 \left(u^4 M_2 +u^3
    \widetilde{M}_2^3\right)\,.
}
\item[-] The Seiberg--Witten curve for the $\mathbb Z_4$ quotient with discrete torsion of the $H_2$ theory is
\al{ y^2 = x^3 -\frac{1}{8} x \left(2 u -M_6\right)^3 M_2 -\frac{1}{8} (2 u -M_6)^4 (u + 2 M_6)\,.
}

\end{itemize}

As an additional comment, we note that here, we have mainly focused on the situation where we treat the $M_i$ as
mass parameters. Of course, since the S-fold introduces a codimension four defect in the worldvolume of the 7-brane,
we can also include additional position dependence in these mass parameters. Doing so would produce F-theory backgrounds
which we can characterize as elliptically fibered Calabi--Yau threefolds in the presence of discrete torsion.

\section{Anomalies}\label{sec:ANOMO}

As a further check on our proposal, in this section we study the scaling of
the conformal anomalies $a$ and $c$ in the limit of large $N$, that is, when
we have a large number of probe D3-branes. We shall also determine the flavor symmetry
anomaly $\kappa_G$ associated with two flavor currents and an R-symmetry current,
namely $\mathrm{Tr}(\mathcal{R} GG)$, where $\mathcal{R}$ denotes the current
for the $U(1)_{\mathcal{R}}$ factor of the R-symmetry $SU(2) \times
U(1)_{\mathcal{R}}$ of a 4D $\mathcal{N} = 2$ SCFT and $G$ refers to a flavor
symmetry current associated with a 7-brane. Since we are dealing with
topological features of the theory,
we will extrapolate our results back to small values of $N$, much as in
reference  \cite{Aharony:2007dj}. From our analysis, we can read off both the
order $N^2$ and order $N$ contributions to the conformal anomalies, however we
will not be able to access the $\mathcal{O}(N^0)$ contributions via these
methods.
This will allow us to compare with the results of reference \cite{Giacomelli:2020jel}, which studies
certain 4D SCFTs from $T^2$ compactifications of 6D $\mathcal{N} = (1,0)$ SCFTs, as well as
with reference \cite{Apruzzi:2020pmv}, which studies some examples of D3-brane probes of
S-folds with discrete torsion. In the rank one case, $N = 1$, we will find
consistency with the rank one theories of \cite{Argyres:2016yzz}, though in those cases we
will have to subtract a free hypermultiplet to match with the interacting
SCFT.

The computation is done using  holography as in \cite{Aharony:2007dj}. The large $N$ dual of the background we are considering is Type IIB on $\text{AdS}_5 \times S^5/\mathbb Z_k$ with 7-branes. We will separate the various terms appearing in the central charges according to their $N$ scaling, with leading order being $N^2$.

\begin{itemize}
\item[-] $\mathcal O(N^2)$: this term comes from the total D3-brane charge induced by the background. The general formula is
\al{ a|_{\mathcal O(N^2)}= c|_{\mathcal O(N^2)} = \frac{M^2 \pi^3}{4 V_5}\,,
}
where $M$ is the D3-brane charge and $V_5$ is the volume of the internal
    five-manifold. In our case $M = N + \varepsilon$ where $\varepsilon =
    \pm(1-k)/2k$ is the charge of the S-fold plane\footnote{Recall that the
    plus sign corresponds to the case without discrete torsion, and the minus
    sign to that with discrete torsion, regardless of the particular choice of
    the discrete torsion.} and $V_5 = \pi^3/k \Delta$.
    The reason for the last identification is that the volume of the
    five-sphere is reduced by a factor of $k$ by the S-fold quotient
    \cite{Aharony:2016kai,Apruzzi:2020pmv} and by a factor of $\Delta$ due to
    the deficit angle of the 7-branes \cite{Aharony:2007dj}.\footnote{$\Delta$
    is both the deficit angle and the dimension of the Coulomb branch
    operator. The values of $\Delta$ are: $\Delta = 6$ for the $E_8$ theory,
    $\Delta = 4$ for the $E_7$ theory, $\Delta = 3$ for the $E_6$ theory,
    $\Delta = 2$ for the $D_4$ theory, $\Delta = 3/2 $ for the $H_2$ theory,
    $\Delta = 4/3$ for the $H_1$ theory and $\Delta = 6/5$ for the $H_0$
    theory.}

\item[-] $\mathcal O(N)$: this term comes from the Chern--Simons terms on the 7-branes. The general formula is
\al{ a|_{\mathcal O(N)} &= \frac{M (\Delta-1)}{2}\,,\\
c|_{\mathcal O(N)} &= \frac{3M (\Delta-1)}{4}\,.
}
As before $M = N + \varepsilon$. Notice that there is no dependence on $k$.
    This is because both the volume wrapped by the 7-branes and the volume of
    the sphere are both affected in the same way by the quotient (the Chern--Simons action is proportional to the ratio of these volumes). Moreover these terms disappear whenever $\Delta = 1$, that is in the case when there are no 7-branes.\footnote{The number of 7-branes is $n_7 = 12 (\Delta-1)/\Delta$.}
%
%
\end{itemize}
{While we have, in principle, been determining the terms at quadratic and
linear orders in $N$, we in fact have determined contributions at
$\mathcal{O}(1)$ from the $\varepsilon$ terms in $M$. We will disregard these
terms, as we cannot determine the $\mathcal{O}(1)$ terms anyway, and we are in
fact required to subtract these terms if the central charges are to match those
occurring for the $\mathcal{N} \geq 3$ theories \cite{Aharony:2016kai}.
Adding the quadratic and linear terms together we get
\al{\label{eqn:ac} a &= \frac{k \Delta}{4} N^2 +  \frac{\left(k \Delta \varepsilon + \Delta-1\right)}{2}  N \,,\\
c &=\frac{k \Delta}{4} N^2 +  \frac{\left( 2 k \Delta \varepsilon + 3 \Delta -3\right)}{4}  N \,.
}
Recall that $\varepsilon = \pm(1-k)/2k$. We can use these formulas and can
check that they agree with the known results for rank one 4D SCFTs
\cite{Argyres:2016yzz}, although in these cases we need to subtract a center
of mass hypermultiplet. In addition, we are able to compute $\kappa_G$, the anomaly
associated with $\mathrm{Tr}(\mathcal{R} GG)$, with $G$ the flavor symmetry generated by the
7-branes in the presence of the S-fold. The results for the cases with
discrete torsion are in \cite{Giacomelli:2020jel}, and here we focus on the
cases without discrete torsion. In general, following \cite{Aharony:2007dj}, one
finds that the central charge for the flavour symmetry $G$ on the 7-branes and
the geometric $SU(2)$ flavour symmetry are
\al{ \kappa_G = 2 N \Delta\,, \quad \kappa_{SU(2)} = kN^2 \Delta -N (\Delta-1-2 k \Delta \varepsilon)\,.
}
Let us note that in the special case where $N = 1$, we always find that either
$\kappa_{SU(2)} = 0$, or that
there is an accidental enhancement in the infrared where the $SU(2)$ merges with the 7-brane flavor symmetry.
We tabulate the values that we get for all cases without discrete torsion writing both the rank $N$ and rank one values,
indicating as well the Kodaira fiber type prior to the quotient. As expected, these are the same values displayed in
reference \cite{Argyres:2016yzz} (for the rank $N$ case the results here match
with \cite{Giacomelli:2020jel}, worked out from compactifications of a 6D
SCFT):
\begin{center}
\begin{tabular}[]{ |c |c |c|c|c|}
\hline
 & $24 a$ & $12 c$& $\kappa_G$& $(24 a,12c,\kappa_G)|_{N=1}$\\
 \hline
$ IV^* / \mathbb Z_2 $
&  $36N^2 +6 N$ & $18N^2 +9N$ & $6N$&(42,27,6)\\
\hline
$ I_0^* / \mathbb Z_2 $
& $24N^2$ & $12 N^2+3N$& $4N$& (24,15,4) \\
\hline
$ IV / \mathbb Z_2 $
& $18N^2-3N$ & $9N^2$ & $3N$ & (15,9,3) \\
\hline
$ I_0^* / \mathbb Z_3 $
&$36 N^2-12 N$  & $18 N^2 -3N$&$4N$& (24,15,4) \\
\hline
$ III / \mathbb Z_3 $
& $24N^2 -12 N$ & $12N^2-5N$&$8N/3$&(12,7,8/3) \\
\hline
$ IV / \mathbb Z_4 $
&  $36N^2-21N$ & $18N^2-9N$&$3N$ &(15,9,3)\\
\hline
\end{tabular}
\end{center}
Here we denoted the theories using the fiber type before taking the quotient
and the type of quotient applied. All the values obtained match with
\cite{Argyres:2016yzz}. Note that the formulas for $a$ and $c$ match the
$\mathcal N=3$ case (obtained when $\Delta = 1$) provided that the $\mathcal
O(1)$ term coming from the center of mass of the system of D3-branes is added
back. For completeness, we can also list the same information in the cases
with discrete torsion, again focusing on the rank one case. As expected, these
are the same values displayed in reference \cite{Argyres:2016yzz} (see also
\cite{Apruzzi:2020pmv,Giacomelli:2020jel}). We can determine these values in
the following manner. We use the formulae in (\ref{eqn:ac}) to
determine the leading and subleading contributions in $N$. The
$\mathcal{O}(1)$ terms were determined in \cite{Apruzzi:2020pmv}, where it was
argued that the parent theory should include $k(\Delta - 1)$ additional free
hypermultiplets before the quotient, and we include them here verbatim.

\begin{center}
\begin{tabular}[]{ |c |c|c|c|c|}
\hline
 & $24 a$ & $12 c$& $\kappa_G$ & $(24 a,12c,\kappa_G)|_{N=1}$ \\
 \hline
$IV^* / \widehat{\mathbb Z}_2 $
& $36N^2+42N+4$ & $18 N^2+27 N+4$ & $6N+1$ & (82,49,7) \\
\hline
$ I_0^* / \widehat{\mathbb Z}_2 $
& $24N^2+24N+2$ &$12N^2 + 15N+2$ & $(4N+1,8N)$ &(50,29,(5,8)) \\
\hline
$ IV / \widehat{\mathbb Z}_2 $
&$18N^2+15+1$ & $9N^2+9N+1$ &$3N+1$ &(34,19,(4,-)) \\
\hline
$ I_0^* / \widehat{\mathbb Z}_3 $
& $36N^2+36N+3$ &$18N^2+21N+3$&$12N+2$&(75,42,14) \\
\hline
$ III / \widehat{\mathbb Z}_3 $
& $24N^2+20N+1$ & $12 N^2 + 11N +1$&-&(45,24,-) \\
\hline
$ IV / \widehat{\mathbb Z}_4 $
&  $36N^2 + 33N+2$ & $18N^2+18N +2$&$12N+2$&(71,38,14) \\
\hline
\end{tabular}
\end{center}
In the above, we have included a ``$-$'' in some entries to reflect
the fact that our present methods do not fix the level of the $U(1)$ flavor current.

\section{Conclusions}\label{sec:CONC}

S-folds are a non-perturbative generalization of O3-planes which figure in the
stringy construction of novel 4D quantum field theories. In this paper we have
proposed a procedure for how S-fold projection acts on the spectrum of string junctions
attached to a stack of 7-branes and probe D3-branes. We have developed a
general prescription for reading off the resulting flavor symmetry algebra
under S-fold projection. This procedure leads to new realizations of many of
the rank one 4D $\mathcal{N} = 2$ SCFTs which arise from mass deformations
and/or discrete gaugings of the rank one $E_8$ Minahan--Nemeschansky theory.
We have also argued that the Seiberg--Witten curves associated with some of
these theories provide an operational definition of F-theory in the presence
of an S-fold background with discrete torsion.  In the remainder of this
section we discuss some avenues for future investigation.

An interesting feature of our analysis is that there is a close correspondence between possible S-fold quotients
of 7-branes, and admissible rank one 4D $\mathcal{N} = 2$ SCFTs. That being said, there are a few examples
which appear in reference \cite{Argyres:2016yzz} which seem to involve some additional ingredients.
The Kodaira fiber types and flavor symmetries
for these cases are $[II^{\ast} , C_2]$, $[III^{\ast} , C_1]$, $[IV^{\ast}_{1}, \varnothing]$, $[II^{\ast} , C_1]$.
In some cases, we can understand the origin of these theories as arising from a mass deformation of another theory, followed by an additional
discrete quotient. That being said, it remains to be understood whether these operations can be fully realized purely in
geometric terms.

There are in principle other ways to generate the same class of rank one 4D $\mathcal{N} = 2$ SCFTs.
In particular, compactifications of 6D SCFTs with suitable discrete twists provide
an alternative way to realize many such examples (see e.g.
\cite{Giacomelli:2020jel}). Since there is now a classification of possible F-theory backgrounds which
can generate 6D SCFTs (see e.g. \cite{Heckman:2013pva, Heckman:2015bfa} and \cite{Heckman:2018jxk} for a review),
it would be interesting to systematically classify all possible ways of incorporating such
discrete effects, thus providing a complementary viewpoint on many of the same questions.

In this paper we have mainly focused on structures associated with 4D $\mathcal{N} = 2$ SCFTs.
It would be quite natural to investigate the structure of related systems with only 4D $\mathcal{N} = 1$ supersymmetry.
For example starting from a 4D $\mathcal{N} = 2$ SCFT, deformations by nilpotent mass deformations often trigger
flows to such theories \cite{Heckman:2010qv, Maruyoshi:2016tqk, Apruzzi:2018xkw}.

O3-planes often play an important role in the construction of consistent Type IIB string vacua.
Having analyzed the effect of S-fold projection on the flavor symmetries of probe D3-branes in the vicinity of
7-branes, it is also natural to consider possible ways in which such ingredients might be used in compact F-theory models.

\section*{Acknowledgments}

We thank M. Del Zotto for helpful discussions.
This work is supported in part by a University Research Foundation
grant at the University of Pennsylvania.

\appendix

\section{Brane Motions}\label{app:BRANE}

In this Appendix we present an illustrative example for how to rearrange various $[p,q]$ 7-branes so that S-fold projection
acts geometrically on the associated string junction states. This is best illustrated via pictures, so we mainly display the relevant figures
here. Our starting point is an $E_6$ stack written as $A^{5} B C^2 \sim A^{6}
X C \sim AAAC AAAC$ (see figure \ref{fig:E6move}), a $D_4$ stack written as $A^{4} BC \sim AAC AAC $ (see figure \ref{fig:D4move})
and an $H_2$ stack written as $A^{3} C \sim AC Y^2 \sim AC AC \sim DADA$ (see figure \ref{fig:H2move}).

\begin{figure}
\begin{center}
\includegraphics[scale = 0.6]{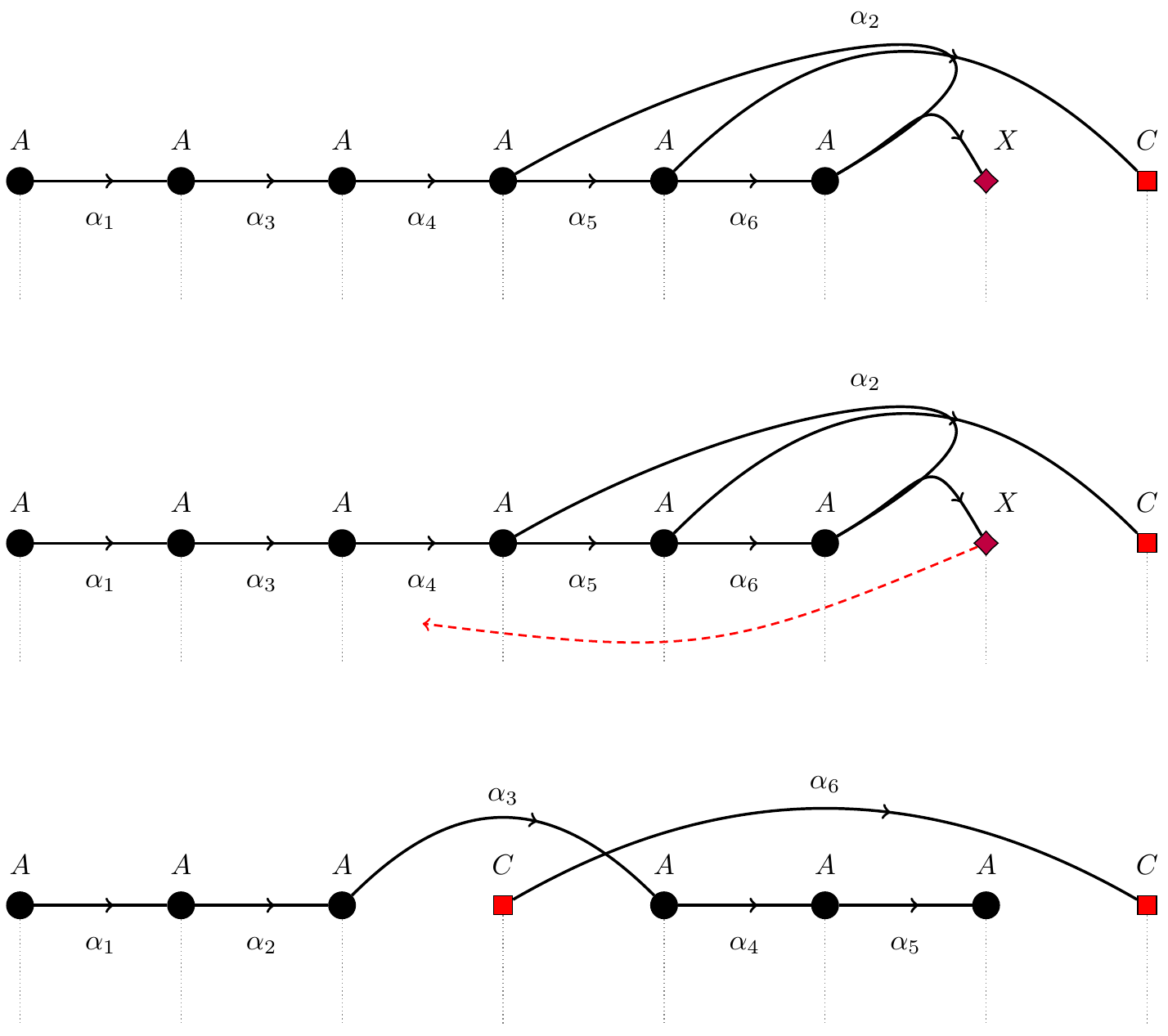}
\end{center}
\caption{Brane motion for $E_6$ 7-branes to a configuration which is $\mathbb{Z}_2$ symmetric, and thus amenable to a $\mathbb{Z}_2$ S-fold projection, i.e. an orientifold projection. In the figure we also indicate how the $X$-brane is moved to accomplish this rearrangement to the $\mathbb{Z}_2$ symmetric configuration $AAACAAAC$.}
\label{fig:E6move}
\end{figure}

\begin{figure}
\begin{center}
\includegraphics[scale = 0.6]{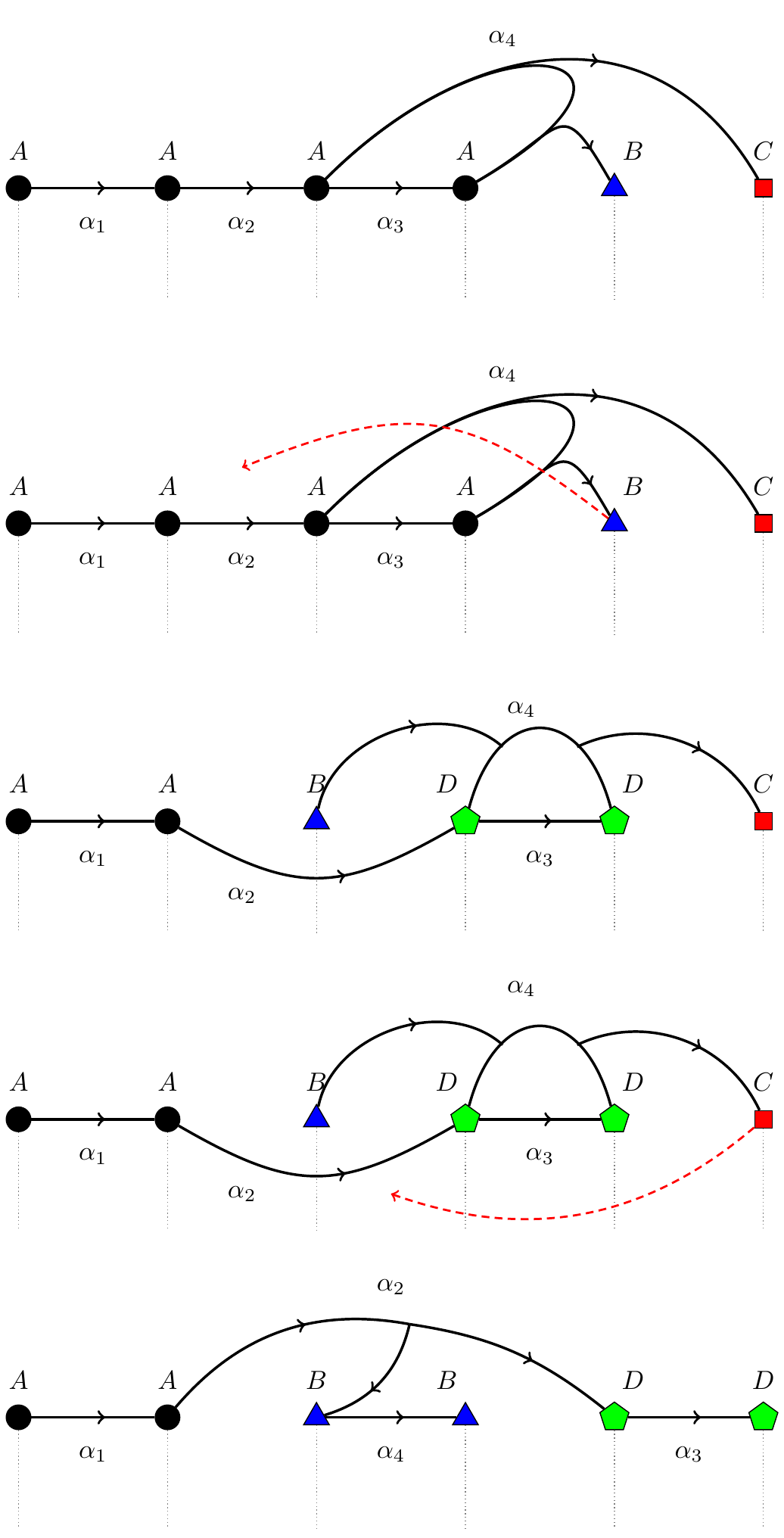}
\end{center}
\caption{Brane motion for $D_4$ 7-branes to a configuration which is $\mathbb{Z}_3$ symmetric, and thus amenable to a $\mathbb{Z}_3$ S-fold projection. In the figure we start with the presentation of this brane system as the bound state $A^{4} B C$, which we then split up into three stacks of branes which are permuted under the $\mathbb{Z}_3$ group action.}
\label{fig:D4move}
\end{figure}

\begin{figure}
\begin{center}
\includegraphics[scale = 0.6]{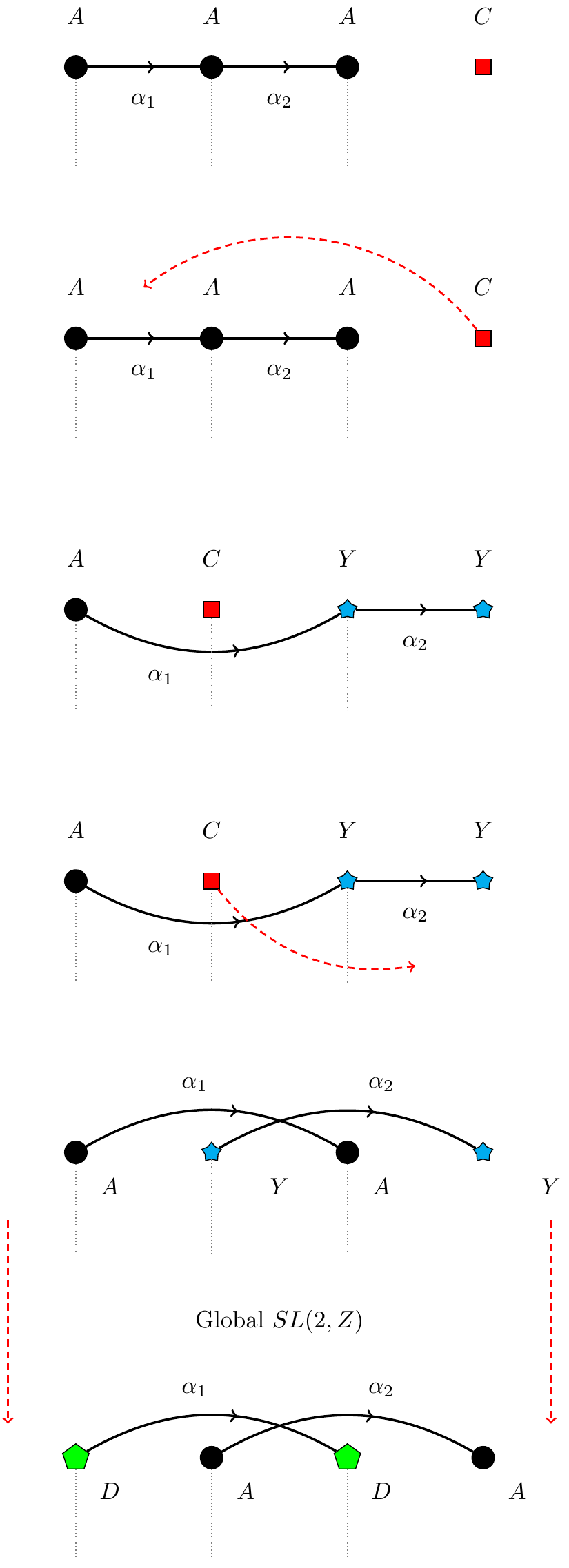}
\end{center}
\caption{Brane motion for $H_2$ 7-branes to the configuration $DADA$ which is $\mathbb{Z}_4$ symmetric, and thus amenable to an S-fold projection. In the last step of rearrangement we apply an $SL(2,\mathbb{Z})$ transformation as indicated by $Y$, with notation as in equation (\ref{monomove}).}
\label{fig:H2move}
\end{figure}

\section{Explicit $\mathbb{Z}_2$ Quotient of $E_6$ without Torsion}\label{app:E6SJ}
In this Appendix we give the explicit root system of $\mathfrak{e}_6$ and show how only $48$ roots survive the $\mathbb{Z}_2$ quotient (without torsion), corresponding exactly to the roots of an $\mathfrak{f}_4$ algebra. The roots of $E_6$, which are given in line \eqref{eq:E6roots} can be written as:
\begin{equation}
  \begin{split}
    \pm\{&(-1, 1, 0, 0, 0, 0, 0, 0), (0, 0, 0, -1, 0, 0, 0, 1), (0, -1, 1, 0, 0, 0, 0, 0), (0, 0, -1, 0, 1, 0, 0, 0),\\
    &(0, 0, 0, 0, -1, 1, 0, 0),(0, 0, 0, 0, 0, -1, 1, 0), (-1, 0, 1, 0, 0, 0, 0, 0), (0, 0, -1, -1, 1, 0, 0, 1),\\
    &(0, -1, 0, 0, 1, 0, 0, 0), (0, 0, -1, 0, 0, 1, 0, 0), (0, 0, 0, 0, -1, 0, 1, 0), (-1, 0, 0, 0, 1, 0, 0, 0),\\
    &(0, -1, 0, -1, 1, 0, 0, 1), (0, 0, -1, -1, 0, 1, 0, 1), (0, -1, 0, 0, 0, 1, 0, 0), (0, 0, -1, 0, 0, 0, 1, 0),\\
    &(-1, 0, 0, -1, 1, 0, 0, 1), (-1, 0, 0, 0, 0, 1, 0, 0), (0, -1, 0, -1, 0, 1, 0, 1), (0, 0, -1, -1, 0, 0, 1, 1), \\
    &(0, -1, 0, 0, 0, 0, 1, 0), (-1, 0, 0, -1, 0, 1, 0, 1), (-1, 0, 0, 0, 0, 0, 1, 0), (0, -1, -1, -1, 1, 1, 0, 1), \\
    &(0, -1, 0, -1, 0, 0, 1, 1), (-1, 0, -1, -1, 1, 1, 0, 1), (-1, 0, 0, -1, 0, 0, 1, 1), (0, -1, -1, -1, 1, 0, 1, 1),\\
    &(-1, -1, 0, -1, 1, 1, 0, 1),  (-1, 0, -1, -1, 1, 0, 1, 1), (0, -1, -1, -1, 0, 1, 1, 1),\\
    &(-1, -1, 0, -1, 1, 0, 1, 1), (-1, 0, -1, -1, 0, 1, 1, 1), (-1, -1, 0, -1, 0, 1, 1, 1),\\
    &(-1, -1, -1, -1, 1, 1, 1, 1), (-1, -1, -1, -2, 1, 1, 1, 2)\},
  \end{split}\label{eq:E6rootvectors}
\end{equation}
where the vectors follow the order of the branes of figure \ref{fig:E6Z2}. Namely, for instance, the highest root: $(1,1,1,2,-1,-1,-1,-2)$ corresponds to the string junction $(a_1+a_2+a_3+2c_1-a_4-a_5-a_6-2c_2)$. For the projection, we define the matrix
\begin{equation}
  Z = -\begin{pmatrix}
    0 & 0 & 0 & 0 & 1 & 0 & 0 & 0 \\
    0 & 0 & 0 & 0 & 0 & 1 & 0 & 0 \\
    0 & 0 & 0 & 0 & 0 & 0 & 1 & 0 \\
    0 & 0 & 0 & 0 & 0 & 0 & 0 & 1 \\
    1 & 0 & 0 & 0 & 0 & 0 & 0 & 0 \\
    0 & 1 & 0 & 0 & 0 & 0 & 0 & 0 \\
    0 & 0 & 1 & 0 & 0 & 0 & 0 & 0 \\
    0 & 0 & 0 & 1 & 0 & 0 & 0 & 0
  \end{pmatrix}.
\end{equation}
We then map every root $r$ in \eqref{eq:E6rootvectors} to $\frac12 (r + Z \cdot r)$. This results in the following $48$ roots:
\begin{equation}
  \begin{split}
    \pm\{
    &(0, 0, 0, -1, 0, 0, 0, 1), (0, 0, -1, 0, 0, 0, 1, 0), (0, -\frac12, \frac12, 0, 0, \frac12, -\frac12, 0),\\
    &(-\frac12, \frac12, 0, 0, \frac12, -\frac12, 0, 0),(0, 0, -1, -1, 0, 0, 1, 1), (0, -\frac12, -\frac12, 0, 0, \frac12, \frac12, 0),\\
    &(-\frac12, 0, \frac12, 0, \frac12, 0, -\frac12, 0),(0, -1, 0, 0, 0, 1, 0, 0), (0, -\frac12, -\frac12, -1, 0, \frac12, \frac12, 1),\\
    &(-\frac12, 0, -\frac12, 0, \frac12, 0, \frac12, 0),(0, -1, 0, -1, 0, 1, 0, 1), (-\frac12, 0, -\frac12, -1, \frac12, 0, \frac12, 1),\\
    &(-\frac12, -\frac12, 0, 0, \frac12, \frac12, 0, 0),(-1, 0, 0, 0, 1, 0, 0, 0), (0, -1, -1, -1, 0, 1, 1, 1),\\
    &(-\frac12, -\frac12, 0, -1, \frac12, \frac12, 0, 1),(-1, 0, 0, -1, 1, 0, 0, 1), (-\frac12, -\frac12, -1, -1, \frac12, \frac12, 1, 1),\\
    &(-1, 0, -1, -1, 1, 0, 1, 1),(-\frac12, -1, -\frac12, -1, \frac12, 1, \frac12, 1), (-1, -\frac12, -\frac12, -1, 1, \frac12, \frac12, 1),\\
    &(-1, -1, 0, -1, 1, 1, 0, 1), (-1, -1, -1, -1, 1, 1, 1, 1), (-1, -1, -1, -2, 1, 1, 1, 2)\}.
  \end{split}\label{eq:F4rootvectors}
\end{equation}
From there, we can extract the four simple roots of $F_4$:
\begin{equation}
  \{(\frac12, -\frac12, 0, 0, -\frac12, \frac12, 0, 0),(0, \frac12, -\frac12, 0, 0, -\frac12, \frac12, 0),(0, 0, 1, 0, 0, 0, -1, 0),(0, 0, 0, 1, 0, 0, 0, -1)\},
\end{equation}
corresponding exactly to the simple roots chosen in line \eqref{eq:F4simpleroots}.

\clearpage

\newpage

\bibliographystyle{utphys}
\bibliography{junct}

\end{document}